\documentclass[aps,prd,groupedaddress,showpacs]{revtex4}

\newcommand{\be}{\begin{equation}}
\newcommand{\ee}{\end{equation}}
\newcommand{\ba}{\begin{eqnarray}}
\newcommand{\ea}{\end{eqnarray}}
\newcommand{\bann}{\begin{eqnarray*}}
\newcommand{\eann}{\end{eqnarray*}}
\newcommand{\nn}{\nonumber}

\newcommand{\intfour}{\int d^4 q\,}
\newcommand{\MeV}{\text{MeV}}
\newcommand{\GeV}{\text{GeV}}
\newcommand{\Trace}{\mbox{\sf Tr}}
\newcommand{\LQCD}{\Lambda_{\text{QCD}}}
\newcommand{\LMSb}{\Lambda_{\overline{\text{MS}}}}

\usepackage{graphicx}
\usepackage{hyperref}

\begin{document}

\preprint{UNITU-THEP-02/02}

\title{{~\hfill \small \rm UNITU-THEP-02/02\\~\\}
Multiplicative renormalizability and quark propagator}

\author{J.C.R. Bloch}
\affiliation{
Institut f\"ur Theoretische Physik, Universit\"at T\"ubingen,
Auf der Morgenstelle 14, D-72076 T\"ubingen, Germany
}

\date{\small February 7, 2002}


\begin{abstract}
The renormalized Dyson-Schwinger equation for the quark propagator is studied, in Landau gauge, in a novel truncation which preserves multiplicative renormalizability. The renormalization constants are formally eliminated from the integral equations, and the running coupling explicitly enters the kernels of the new equations. To construct a truncation which preserves multiplicative renormalizability, and reproduces the correct leading order perturbative behavior, non-trivial cancellations involving the full quark-gluon vertex are assumed in the quark self-energy loop. A model for the running coupling is introduced, with infrared fixed point in agreement with previous Dyson-Schwinger studies of the gauge sector, and with correct logarithmic tail. Dynamical chiral symmetry breaking is investigated, and the generated quark mass is of the order of the extension of the infrared plateau of the coupling, and about three times larger than in the Abelian approximation, which violates multiplicative renormalizability. The generated scale is of the right size for hadronic phenomenology, without requiring an infrared enhancement of the running coupling.
\end{abstract}

\maketitle

\section{Introduction}

In the Standard Model of the strong, weak and electromagnetic forces, the interactions are quantitatively described by gauge field theories. Quantum chromodynamics is a non-Abelian gauge theory, and the proof of its renormalizability \cite{'tHooft:1971fh} and discovery of ultraviolet asymptotic freedom \cite{Politzer:1973fx} have been milestones in its acceptance as theory of the strong interaction. For large momenta the coupling becomes small, and perturbation theory seems an appropriate calculational tool. However, for small momenta the coupling grows large and adequate methods have to be used to study nonperturbative phenomena like confinement, dynamical chiral symmetry breaking and bound state formation. One such method is the study of the Dyson-Schwinger equations (DSE) \cite{Roberts:1994dr}, and their phenomenological applications to hadronic physics is a subject of growing interest \cite{Roberts:2000aa}. 

Dynamical chiral symmetry breaking can be studied by means of the quark propagator DSE, also called gap equation. The quark equation is part of an infinite tower of integral equations relating all the Green's functions of the quantum field theory, and a truncation is necessary to be able to solve it. The often used Abelian approximation introduces an effective running coupling in the kernels of the quark equation, based on features of QED. A number of studies of the gap equation were performed in this approximation with effective strong couplings that are infrared enhanced \cite{Brown:1988bm, Frank:1996uk, Maris:1997tm}, infrared vanishing \cite{Hawes:1994ef, Hawes:1998cw}, infrared vanishing with large enhancements in the intermediate region \cite{Jain:1993qh}, and infrared finite \cite{Atkinson:1988mw, Roberts:1990mj}. These studies all find that dynamical chiral symmetry breaking can be triggered, provided some parameter in the model exceeds a critical value, and a large integration strength is needed in the self-energy kernels to achieve large enough "constituent" quark masses, needed for hadron phenomenology \cite{Hawes:1998cw}. 
The Ansatz of a strong infrared enhancement in the kernels of the quark DSE is therefore a major ingredient in phenomenological studies, and the search for the source of such an enhancement, coming from the gluon propagator, quark-gluon vertex or their combination is an important field of investigation. 

The kernel of the gap equation not only reflects the strong interaction between quarks and gluons, but also embodies the complicated structure of the QCD vacuum, which in the continuum contains self-interacting gluon fields as well as ghost fields. Understanding the infrared behavior of the gluon and ghost propagators, and of the running coupling is therefore of great importance for the study of dynamical mass generation \cite{Alkofer:2000wg}.

Early studies of the Dyson-Schwinger equation for the gluon propagator in the Landau gauge seemed to indicate that the gluon propagator could be highly singular in the infrared \cite{Mandelstam:1979xd, Atkinson:1981er, Atkinson:1982ah, Brown:1989bn}, possibly providing the above mentioned infrared enhancement in the kernels of the gap equation.
However, these gluon propagator studies neglected any contribution of the ghost fields, and required the ad hoc cancellations of certain leading terms in the equations. It is therefore far from certain that these solutions reflect the correct QCD infrared behavior.

More recently, studies of the coupled set of Dyson-Schwinger equations for the gluon and ghost propagators \cite{vonSmekal:1998is, Atkinson:1998tu, Atkinson:1998zc} have shown that the coupling of ghost and gluon fields plays a crucial role in the generation of a consistent infrared behavior of QCD. In the infrared the gluon and ghost propagators are power behaved, and the strong running coupling has an infrared fixed point. Moreover, the leading infrared power in the gluon vacuum polarization depends only on the ghost loop, and not on the gluon loop nor any other diagram in the gluon equation. Although the precise values of the infrared powers depend on the vertex Ans\"atze and other details of the truncation, all these studies showed that the full ghost propagator is more divergent than its bare counterpart, while the full gluon propagator is less divergent. Recent lattice calculations also support the infrared propagator power laws and fixed point of the running coupling \cite{Bonnet:2000kw, Langfeld:2001cz, Langfeld:2001a}. As neither the gluon propagator, nor the running coupling are infrared enhanced in the Landau gauge, one seems to be at loss to explain how enough strength can be achieved when breaking the chiral symmetry in the gap equation. The problem to reconcile the DSE results of the gauge sector with the need for a strong infrared enhancement in the effective coupling might hint to a deficiency in the Abelian approximation to the gap equation. A first attempt to extend the study of dynamical chiral symmetry breaking beyond the Abelian approximation, by solving the coupled set of DSEs for the quark, gluon and ghost propagators simultaneously \cite{Ahlig:thesis}, underlines this conundrum by finding a generated fermion scale that is a factor two too small for phenomenological purposes. One can then wonder if the approximations, introduced to solve the gap equation, violate some essential property of quantum chromodynamics, such that the strength of dynamical chiral symmetry breaking is significantly weakened? And if so, can we construct a truncation which preserves this property, and would naturally generate a large enough dynamical quark mass?

Multiplicative renormalizability (MR) is an important feature of gauge field theories, proven for perturbative renormalization \cite{'tHooft:1971fh}, but believed to hold for \textit{any} renormalization scale, and it is often ruined by the approximations to the vertices and vertex renormalization constants introduced in DSE studies to truncate the infinite tower of equations. None of the truncations introduced in the DSE studies mentioned above, on the quark, gluon and ghost DSEs, preserve multiplicative renormalizability. 

In a recent paper \cite{Bloch:2001wz} we have discussed how the DSE studies \cite{vonSmekal:1998is, Atkinson:1998tu, Atkinson:1998zc} of the gauge sector violate the multiplicative renormalizability of the gluon and ghost propagators, and we developed a new truncation scheme which preserves multiplicative renormalizability, and showed how this induced important changes in the equations. 
We investigated the consequences for the analysis of the infrared behavior of the propagators, and
found that the ghost loop, gluon loop, 3-gluon and 4-gluon diagrams all \textit{potentially} contribute to the leading infrared power in the gluon propagator DSE, which is at variance with the conclusions of Refs. \cite{vonSmekal:1998is, Atkinson:1998tu, Atkinson:1998zc}, where only the ghost loop contributed to leading order. 
Although the combination of ghost and gluon loops alone does not seem to allow infrared power solutions for the propagators \cite{Bloch:2001wz}, there are strong indications that the infrared contribution of the 4-gluon diagram, also called squint diagram, is such that infrared power solutions do exist when all the diagrams are taken into account \cite{Bloch:2001b}.

In this paper we construct an MR preserving truncation to the quark DSE using an approach similar to that presented in Ref.\ \cite{Bloch:2001wz} for the gauge sector, and investigate 
dynamical chiral symmetry breaking. 
We show that the preservation of multiplicative renormalizability in the truncated quark equation can naturally generate a phenomenologically acceptable mass scale, while being consistent with the properties of the gluon and ghost propagators known from the DSE studies of the gauge sector, without infrared enhancement of the strong running coupling. 

We first reformulate the quark equations, such that the Landau gauge equations are free of renormalization constants. In this formulation the self-energy kernels are explicitly driven by the running coupling, and the multiplicative renormalizability of the quark dressing function, and renormalization point invariance of the dynamical mass are manifest. This allows us to design a novel truncation scheme for these formal equations, which respects both the multiplicative renormalizability and the resummed perturbative limit of the solutions. The full vertex is genuinely nonperturbative in our truncation, and non-trivial cancellations in the self-energy integrals are assumed to strip the vertex from its complete dressing, leaving only bare vertices in the truncated equations. 

In the new truncated quark equations, the running coupling is the only object carrying information about the nonperturbative effects in the gauge sector. The back reaction of the quarks on the running coupling could in principle be included through a self-consistent study of the coupled quark-gluon-ghost DSEs. However, to perform a first quantitative calculation of dynamical chiral symmetry breaking, we introduce a model running coupling based on prior knowledge about its infrared fixed point, acquired from the ghost-gluon DSE \cite{Atkinson:1998zc, Bloch:2001wz}, the resummed leading order perturbative behavior, and the qualitative features of the transition between both regions \cite{Atkinson:1998tu}.

Dynamical chiral symmetry breaking is then investigated using this model running coupling in the MR truncation of the quark DSE. Although the equations in our new truncation look very similar to those in the Abelian approximation, the slight modifications ensuring MR yield important quantitative differences: the dynamically generated mass in the MR truncation is boosted by about a factor three compared to the Abelian approximation with same running coupling. Our new findings show that no infrared enhancement, nor fine-tuning of the strong running coupling is needed: a running coupling having a reasonably valued infrared fixed point ($\alpha_0 \approx 2.6$), which smoothly bends over into the perturbative logarithmic tail, is able to generate masses allowing for realistic hadron phenomenology.

\section{The quark equation}

Starting from the spinor quark DSE \cite{Roberts:1994dr}, we derive two scalar equations for the mass function and renormalized quark dressing function, using the definitions of renormalized fields and coupling. The ensuing quark equations are part of an infinite tower of integral equations relating all the Green's functions of QCD, and a truncation is necessary to be able to solve them. We briefly discuss the Abelian approximation, which violates multiplicative renormalizability, before constructing a novel truncation which preserves the multiplicative renormalizability of the quark solutions.  

The Dyson-Schwinger equation for the quark propagator in QCD, formulated in Euclidean space, is
\be\label{quark-dse}
[S_F(p)]^{-1} = [S_F^0(p)]^{-1}
- C_F \, g_0^2 \int \frac{d^4 q}{(2\pi)^4} \, \Gamma^{qg,0}_\mu(p,q,r) \, S_F(q) \, \Gamma^{qg}_\nu(q,p,-r) \, D^{\mu\nu}(r)  \,,
\ee
where $S_F$ and $S_F^0$ are the full and bare quark propagators, $D^{\mu\nu}$ is the full gluon propagator, $\Gamma^{qg,0}_\mu$ and $\Gamma^{qg}_\nu$ are the bare and full quark-gluon vertices, the color factor $C_F=(N_c^2-1)/2N_c = 4/3$ for $N_c=3$, $g_0$ is the bare coupling, and $r=p-q$.

The most general expression for the full quark propagator can be written as
\be\label{qprop}
S_F(p) = \frac{Z(p^2)} { i p \cdot \gamma + M(p^2)} \,,
\ee
where $Z(p^2)$ is the quark dressing function, and $M(p^2)$ is the mass function.
The bare propagator for a quark with bare mass $m_0$ is
\be\label{qprop0}
S_F^0(p) = \frac{1} { i p \cdot \gamma + m_0} \,.
\ee
The full gluon propagator in a general covariant gauge is given by
\be\label{glprop}
D_{\mu\nu}(p) =  \left(\delta_{\mu\nu}-\frac{p_\mu p_\nu}{p^2}\right) \frac{F(p^2)}{p^2} + \xi \frac{p_\mu p_\nu}{p^4} \,,
\ee
where $F(p^2)$ is the gluon dressing function, and for future use we also give the general form of the full ghost propagator:
\be\label{ghprop}
\Delta(p) = - \frac{G(p^2)}{p^2} \,,
\ee
where $G(p^2)$ is the ghost dressing function.

After substituting the propagators (\ref{qprop}) - (\ref{glprop}) in Eq.\ (\ref{quark-dse}), we can derive a set of two coupled equations for the quark dressing function and the mass function by multiplying Eq.\ (\ref{quark-dse}) respectively by $- i p\cdot\gamma$ and $I_4$, and taking the trace in spinor space:
\ba\label{Zx}
\frac{1}{Z(p^2,\Lambda^2)} &=& 1 + \frac{C_F}{16\pi^4}  \, g_0^2(\Lambda^2) \intfour  \, Z(q^2,\Lambda^2) \, \frac{F(r^2,\Lambda^2)}{q^2 + M^2(q^2)} \, U_Z(p^2,q^2,r^2,\Lambda^2) \,,\\
\label{Mx}
\frac{M(p^2)}{Z(p^2,\Lambda^2)} &=& m_0(\Lambda^2) - \frac{C_F}{16\pi^4}  \, g_0^2(\Lambda^2)  \intfour \, Z(q^2,\Lambda^2) \, M(q^2) \, \frac{F(r^2,\Lambda^2)}{q^2 + M^2(q^2)} \, U_M(p^2,q^2,r^2,\Lambda^2) \,,
\ea
where 
\ba
\label{UZ}
U_Z(p^2,q^2,r^2,\Lambda^2) &=& \frac{1}{4 p^2 r^2} \Trace\bigg[p\cdot\gamma \, \Gamma_\mu^{qg,0}(p,q,r) \, \left(q\cdot\gamma + i \, M(q^2)\right) \, \Gamma_\nu^{qg}(q,p,-r,\Lambda^2)\bigg]\left[\delta_\perp^{\mu\nu}(r) + \frac{\xi}{ F(r^2,\Lambda^2)}\frac{r^\mu r^\nu}{r^2}\right] ,\\
\label{UM}
U_M(p^2,q^2,r^2,\Lambda^2) &=& \frac{1}{4 r^2} \Trace\bigg[\Gamma_\mu^{qg,0}(p,q,r) \, \left(1 - i \, \frac{q\cdot\gamma}{M(q^2)}\right) \, \Gamma_\nu^{qg}(q,p,-r,\Lambda^2)\bigg]\left[\delta_\perp^{\mu\nu}(r) + \frac{\xi}{ F(r^2,\Lambda^2)}\frac{r^\mu r^\nu}{r^2}\right] \,,
\ea
with $\delta_\perp^{\mu\nu}(r) = g^{\mu\nu} - r^\mu r^\nu/r^2$, and the kernels $U_Z$ and $U_M$ also depend implicitly on $Z$ and $M$ through the full quark-gluon vertex $\Gamma_\nu^{qg}$. The vector and scalar quark self-energy integrals are herein regulated with an ultraviolet cutoff $\Lambda$, yielding a $\Lambda^2$ dependence of the regularized dressing functions, and the $\Lambda^2$ dependence of the bare mass $m_0(\Lambda^2)$ is such that $M(p^2)$ is finite and independent of $\Lambda^2$.

\subsection*{Renormalization}

The full, regularized Green's functions are potentially divergent when the ultraviolet cutoff $\Lambda$ is taken to infinity, and they are renormalized by applying the principles of multiplicative renormalization.
The quark field is renormalized by
\be\label{renormZ}
Z(p^2, \Lambda^2) = Z_2(\mu^2,\Lambda^2) \, Z_R(p^2, \mu^2) \,,
\ee
where $Z_R$ is the renormalized quark dressing function, $Z_2$ is the quark field renormalization constant, and at the renormalization point $Z_R(\mu^2, \mu^2) \equiv 1$. 

When addressing mass renormalization it is important to observe that each choice of bare mass parameter $m_0(\Lambda^2)$ defines a different physical theory \cite{Collins:1984xc}, and for QCD each quark flavor corresponds to a different value $m^q_0(\Lambda^2)$. The running mass function $M(p^2)$ is renormalization point invariant as it is unambiguously determined by $m_0(\Lambda^2)$. The bare mass parameter is chosen such that the running mass function $M(p^2)$ is finite, and takes the value $M(\mu^2)=m_\mu$, where the renormalized mass parameter $m_\mu$ is determined, in practice, by matching the calculated value of some mass dependent observable to its measured value. 
Note that we speak of mass \textit{parameter} instead of mass, to emphasize that the quarks are confined and have therefore no pole mass. 

The renormalization of the quark-gluon vertex introduces a vertex renormalization constant $Z_{1f}$, which is related to the renormalization of the strong coupling by
\be\label{renormq}
g(\mu^2) = \frac{Z_3^{1/2}(\mu^2,\Lambda^2)
Z_2(\mu^2,\Lambda^2)}{Z_{1f}(\mu^2,\Lambda^2)} \, g_0(\Lambda^2) \,,
\ee
where $g(\mu^2)$ is the value of the renormalized coupling at the renormalization scale. Gauge invariance of the renormalized theory ensures the universality of the renormalized coupling for the quark-gluon, ghost-gluon and triple-gluon interaction, such that also 
\be\label{renormghgl}
g(\mu^2) = \frac{Z_3^{1/2}(\mu^2,\Lambda^2)\tilde
Z_3(\mu^2,\Lambda^2)}{\tilde{Z}_1(\mu^2,\Lambda^2)} \, g_0(\Lambda^2)
= \frac{Z_3^{3/2}(\mu^2,\Lambda^2)}{Z_1(\mu^2,\Lambda^2)} \, g_0(\Lambda^2)\,, 
\ee
where $\tilde Z_1$ and $Z_1$ are the ghost-gluon and triple-gluon vertex renormalization constants. $Z_3$ and $\tilde Z_3$ are the renormalization constants for the gluon and ghost fields, 
\be\label{renorm}
F(p^2, \Lambda^2) = Z_3(\mu^2,\Lambda^2) \, F_R(p^2, \mu^2) \quad , \qquad
G(p^2, \Lambda^2) = \tilde Z_3(\mu^2,\Lambda^2) \, G_R(p^2, \mu^2)  \,,
\ee
where the renormalized gluon and ghost dressing functions are defined such that $F_R(\mu^2, \mu^2)=G_R(\mu^2, \mu^2) \equiv 1$.

Now that all the necessary renormalized quantities have been defined, we multiply Eqs.\ (\ref{Zx}), (\ref{Mx}) by $Z_2$, such that their left hand sides become finite, and introduce the renormalized dressing functions using Eqs.\ (\ref{renormZ}), (\ref{renorm}), and the renormalized coupling $g(\mu^2)$ using Eq.\ (\ref{renormq}):
\ba\label{Zxb}
\frac{1}{Z_R(p^2,\mu^2)} &=& Z_2(\mu^2,\Lambda^2) + \frac{C_F }{4\pi^3}  \, \alpha(\mu^2) \, Z_{1f}^2(\mu^2,\Lambda^2) \, \intfour \, Z_R(q^2,\mu^2) \, \frac{F_R(r^2,\mu^2)}{q^2 + M^2(q^2)} \, U_Z(p^2,q^2,r^2,\Lambda^2) \, , \\
\frac{M(p^2)}{Z_R(p^2,\mu^2)} &=& Z_2(\mu^2,\Lambda^2) \, m_0(\Lambda^2) 
- \frac{C_F}{4\pi^3}  \, \alpha(\mu^2) \, Z_{1f}^2(\mu^2,\Lambda^2) 
\intfour \, Z_R(q^2,\mu^2) \, M(q^2) \, \frac{F_R(r^2,\mu^2)}{q^2 + M^2(q^2)} \, U_M(p^2,q^2,r^2,\Lambda^2) \, , \nn\\
\label{Mxb}
\ea
where we defined $\alpha(\mu^2) \equiv g^2(\mu^2)/4\pi$. Note that the integrals in the renormalized equations are preceded by a renormalization factor $Z_{1f}^2(\mu^2,\Lambda^2)$, originating from the renormalization of the coupling. Although it is customary to absorb one $Z_1$ factor in the full unrenormalized vertex in order to renormalize it, we deliberately do not proceed like this here, so that the full vertex depends on the momenta flowing in the vertex, and on $\Lambda$, but not on $\mu$. The self-energy integrals have a $\Lambda$ dependence consistent with perturbation theory, which is cancelled by the seeds of Eqs.\ (\ref{Zxb}), (\ref{Mxb}) to yield a finite dressing function $Z_R$ and mass function $M$.

\subsection*{Abelian approximation}

A truncation often used in previous studies to decouple the quark equations (\ref{Zxb}), (\ref{Mxb}) from the infinite tower of DSEs is the bare vertex, Abelian approximation.
There, ghost field contributions are neglected, and one introduces the substitution
\begin{equation}\label{Abelian}
\alpha(\mu^2) \, Z_{1f}^2 \, F_R(r^2,\mu^2) \, \Gamma^{qg,0}_\mu \, \Gamma^{qg}_\nu
\longrightarrow \alpha^{\text{eff}}(r^2) \, \Gamma^{qg,0}_\mu \, \Gamma^{qg,0}_\nu \,,
\end{equation}
based on QED gauge invariance properties, and on the assumption that the bare vertex approximation might be appropriate in the Landau gauge. In this approximation $\alpha^{\text{eff}}(r^2)$ can be considered as an \textit{effective} running coupling. 
In the Abelian approximation the quark equations (\ref{Zxb}), (\ref{Mxb}) become
\ba\label{ZxAb}
\frac{1}{Z_R(p^2,\mu^2)} &=& Z_2(\mu^2,\Lambda^2) + \frac{C_F }{4\pi^3}  \, \intfour \, Z_R(q^2,\mu^2) \, \frac{\alpha^{\text{eff}}(r^2)}{q^2 + M^2(q^2)} \, U_Z^0(p^2,q^2,r^2) \,,\\
\label{MxAb}
\frac{M(p^2)}{Z_R(p^2,\mu^2)} &=& Z_2(\mu^2,\Lambda^2) \, m_0(\Lambda^2) 
- \frac{C_F}{4\pi^3} \, \intfour \, Z_R(q^2,\mu^2) \, \, M(q^2) \frac{\alpha^{\text{eff}}(r^2)}{q^2 + M^2(q^2)} \, U_M^0(p^2,q^2,r^2) \,,
\ea
where $U_Z^0$ and $U_M^0$ are calculated by replacing the full, regularized, unrenormalized quark-gluon vertex $\Gamma^{qg}_\nu$ by the bare vertex, $\Gamma^{qg,0}_\nu = i \gamma_\nu$, in Eqs.\ (\ref{UZ}), (\ref{UM}).

It is easy to convince oneself that Eqs.\ (\ref{ZxAb}), (\ref{MxAb}) of the Abelian approximation violate multiplicative renormalizability, and this will be illustrated further in the result section.

\subsection*{MR truncation}

The aim of this paper is to derive a novel truncation scheme, which preserves the multiplicative renormalizability of the solutions, starting from the exact renormalized equations (\ref{Zxb}), (\ref{Mxb}). This will be achieved by specific manipulations of the renormalization constant $Z_{1f}$, and by assuming non-trivial cancellations involving the full quark-gluon vertex in the quark self-energy loop, as specified in the remaining of this section.

The nonperturbative expression for the quark-gluon vertex renormalization constant $Z_{1f}$ is not known, but we can eliminate it from our equations using  Eqs.\ (\ref{renormq}), (\ref{renormghgl}), which state the universality of the strong coupling:
\be\label{Z1f}
Z_{1f}(\mu^2,\Lambda^2) = \frac{Z_2(\mu^2,\Lambda^2)}{\tilde Z_3(\mu^2,\Lambda^2)} \, \tilde{Z}_1(\mu^2,\Lambda^2)\,.
\ee
It is interesting to note the similarity between the QCD relation (\ref{Z1f}) and the QED analog, $Z_{1f} = Z_2$, in a theory without ghost fields. 

As $Z_{1f}$ is independent of the integration momentum, the factor $Z_{1f}^2$ can be moved inside the self-energy integrals of Eqs.\ (\ref{Zxb}), (\ref{Mxb}), and factorized in a $\Lambda$ and a $\mu$ dependent part by applying Eq.\ (\ref{Z1f}), and consequently using the definitions (\ref{renormZ}), (\ref{renorm}) to replace $Z_2$ and $\tilde Z_3$ by ratios of unrenormalized to renormalized dressing functions at momenta of our choice: 
\be\label{elimZ1f}
Z_{1f}(\mu^2,\Lambda^2)
= \frac{Z(q^2,\Lambda^2)}{G(r^2,\Lambda^2)} \frac{G_R(r^2,\mu^2)}{Z_R(q^2,\mu^2)}  \, \tilde{Z}_1(\mu^2,\Lambda^2) \,,
\ee
where we have chosen to use the quark momentum of the self-energy loop to rewrite $Z_2$, and the gluon momentum for $\tilde Z_3$. After substituting the square of Eq.\ (\ref{elimZ1f}) in the quark equations (\ref{Zxb}), (\ref{Mxb}) we find
\ba
\frac{1}{Z_R(p^2,\mu^2)} &=& Z_2(\mu^2,\Lambda^2) + \frac{C_F}{4\pi^3} \alpha(\mu^2) \tilde{Z}_1^2(\mu^2,\Lambda^2) \nn\\
\label{Zxd}
&& \times\intfour \, \frac{1}{Z_R(q^2,\mu^2)} \, \frac{F_R(r^2,\mu^2) \, G_R^2(r^2,\mu^2)}{q^2 + M^2(q^2)}  \, \left[\frac{Z^2(q^2,\Lambda^2)}{G^2(r^2,\Lambda^2)}\, U_Z(p^2,q^2,r^2,\Lambda^2)\right] \,,\\
\frac{M(p^2)}{Z_R(p^2,\mu^2)} &=& Z_2(\mu^2,\Lambda^2) \, m_0(\Lambda^2) 
- \frac{C_F}{4\pi^3} \alpha(\mu^2) \tilde{Z}_1^2(\mu^2,\Lambda^2) \nn\\
\label{Mxd}
&& \times\intfour \, \frac{M(q^2)}{Z_R(q^2,\mu^2)} \, \frac{F_R(r^2,\mu^2) \, G_R^2(r^2,\mu^2)}{q^2 + M^2(q^2)}  \, \left[\frac{Z^2(q^2,\Lambda^2)}{G^2(r^2,\Lambda^2)} \, U_M(p^2,q^2,r^2,\Lambda^2) \right] \,.
\ea

From Eqs.\ (\ref{renormghgl}), (\ref{renorm}) it is easy to see that the product 
\be
\label{rgi2}
\hat\alpha(q^2,\mu^2,\Lambda^2) \equiv  \alpha(\mu^2) \, \tilde Z_1^2(\mu^2,\Lambda^2) \, F_R(q^2,\mu^2) \, G_R^2(q^2,\mu^2)
\ee
can also be written in terms of unrenormalized quantities only:
\be
\label{rgi3}
\hat\alpha(q^2,\mu^2,\Lambda^2) = \frac{g^2_0(\Lambda^2)}{4\pi} \, F(q^2,\Lambda^2) \, G^2(q^2,\Lambda^2) \, ,
\ee
which shows that $\hat\alpha(q^2,\mu^2,\Lambda^2)$ is renormalization group invariant, i.e.\ independent of $\mu^2$. Therefore, $\hat\alpha(q^2,\mu^2,\Lambda^2) = \hat\alpha(q^2,q^2,\Lambda^2)$, and taking into account the renormalization conditions for $F_R$ and $G_R$, and the fact that $\tilde Z_1 \equiv 1$ in the Landau gauge \cite{Taylor:1971ff}, we see that, in that gauge, the product $\hat\alpha(q^2,\mu^2,\Lambda^2)$ defined in Eq.\ (\ref{rgi2}) is nothing else but the running coupling $\alpha(q^2)$ (see also Refs.\ \cite{Mandelstam:1979xd, vonSmekal:1998is, Atkinson:1998tu}).
One can identify the product (\ref{rgi2}) in Eqs.\ (\ref{Zxd}), (\ref{Mxd}), and in the Landau gauge these equations become:
\ba\label{Zxe}
\frac{1}{Z_R(p^2,\mu^2)} &=& Z_2(\mu^2,\Lambda^2) + \frac{C_F}{4\pi^3} 
\intfour \, \frac{1}{Z_R(q^2,\mu^2)} \, \frac{\alpha(r^2)}{q^2 + M^2(q^2)}  \, \left[\frac{Z^2(q^2,\Lambda^2)}{G^2(r^2,\Lambda^2)}\, U_Z(p^2,q^2,r^2,\Lambda^2)\right] \,,\\
\label{Mxe}
\frac{M(p^2)}{Z_R(p^2,\mu^2)} &=& Z_2(\mu^2,\Lambda^2) \, m_0(\Lambda^2) 
- \frac{C_F}{4\pi^3} 
\intfour \, \frac{M(q^2)}{Z_R(q^2,\mu^2)} \, \frac{\alpha(r^2)}{q^2 + M^2(q^2)}  \, \left[\frac{Z^2(q^2,\Lambda^2)}{G^2(r^2,\Lambda^2)} \, U_M(p^2,q^2,r^2,\Lambda^2) \right] \,.
\ea
In numerical calculations the seeds of the equations will be eliminated in the usual way by subtracting each equation at two different momenta, and imposing the renormalization conditions. Note that, except for the seeds of the integral equations (\ref{Zxe}), (\ref{Mxe}), the renormalization point dependence only enters through the renormalized quark dressing function $Z_R$, so that the multiplicative renormalizability of the dressing function, and renormalization point invariance of the mass function are manifested in a very clear way by the $1/Z_R$ structure on both sides of the equations. Furthermore these equations are still exact as no approximations have been introduced in their derivation, and we have shown how the running coupling explicitly enters the self-energy kernels, and drives the mass function, in a Non-Abelian gauge theory. 

The kernels $U_Z$ and $U_M$, defined in Eqs.\ (\ref{UZ}), (\ref{UM}), contain the full, unrenormalized quark-gluon vertex, thus coupling the propagator equations (\ref{Zxe}), (\ref{Mxe}) to the higher order DSEs, and in order to perform a numerical calculation we introduce an approximation which decouples the quark equations from the vertex DSE. 

A simple truncation which preserves the MR properties of the solutions, and reproduces the leading order renormalization group equation (RGE) improved perturbative behavior consists in assuming that the factors $Z^2/G^2$ in the square brackets of Eqs.\ (\ref{Zxe}), (\ref{Mxe}) cancel both the nonperturbative vertex and loop corrections in the self-energy integrals. This truncation is analogous to the one introduced in the DSE for the gluon propagator in Ref.\ \cite{Bloch:2001wz}. The kernels between square brackets in Eqs.\ (\ref{Zxe}), (\ref{Mxe}) are thus approximated by
\ba
\label{UZ0}
\frac{Z^2(q^2,\Lambda^2)}{G^2(r^2,\Lambda^2)} \, U_Z(p^2,q^2,r^2,\Lambda^2) &\longrightarrow& U_Z^0(p^2,q^2,r^2) \\
\label{UM0}
\frac{Z^2(q^2,\Lambda^2)}{G^2(r^2,\Lambda^2)} \, U_M(p^2,q^2,r^2,\Lambda^2) &\longrightarrow& U_M^0(p^2,q^2,r^2) \,,
\ea
where $U_Z^0$ and $U_M^0$ are calculated  by replacing the full, regularized, unrenormalized quark-gluon vertex $\Gamma^{qg}_\nu$ by the bare vertex, $\Gamma^{qg,0}_\nu = i \gamma_\nu$, in Eqs.\ (\ref{UZ}), (\ref{UM}). After substituting the Ans\"atze (\ref{UZ0}), (\ref{UM0}) in Eqs.\ (\ref{Zxe}), (\ref{Mxe}) we find \ba\label{Zxf}
\frac{1}{Z_R(p^2,\mu^2)} &=& Z_2(\mu^2,\Lambda^2) + \frac{C_F}{4\pi^3} 
\intfour \, 
  \frac{1}{Z_R(q^2,\mu^2)} \, \frac{\alpha(r^2)}{q^2 + M^2(q^2)}  \, U_Z^0(p^2,q^2,r^2)\,,\\
\label{Mxf}
\frac{M(p^2)}{Z_R(p^2,\mu^2)} &=& Z_2(\mu^2,\Lambda^2) \, m_0(\Lambda^2) - \frac{C_F}{4\pi^3} 
\intfour \,
\frac{M(q^2)}{Z_R(q^2,\mu^2)} \, \frac{\alpha(r^2)}{q^2 + M^2(q^2)} \, U_M^0(p^2,q^2,r^2) \,,
\ea
where the Landau gauge kernels ($\xi = 0$) are given by
\ba\label{UZ1}
U_Z^0(p^2,q^2,r^2) &=& \frac{1}{p^2 r^2} \left[ 3 p \cdot q - \frac{2 \,( p^2 q^2 - (p \cdot q)^2)}{r^2} \right] \,, \\
\label{UM1}
U_M^0(p^2,q^2,r^2) &=& - \frac{3}{r^2}  \,.
\ea

Let us verify that this truncation preserves the multiplicative renormalizability of the solutions. From Eq.\ (\ref{renormZ}) one can show that multiplicative renormalization is satisfied if solutions renormalized at $\mu^2$ and $\nu^2$ are related by:
\be\label{ZRnu}
Z_R(p^2,\nu^2) = \frac{Z_R(p^2,\mu^2)}{Z_R(\nu^2,\mu^2)} \quad , \qquad
Z_2(\nu^2, \Lambda^2) = Z_R(\nu^2,\mu^2) \, Z_2(\mu^2, \Lambda^2) \,.
\ee
Now assume one has found solutions $Z_R(p^2,\mu^2)$, $M(p^2)$ of Eqs.\ (\ref{Zxf}), (\ref{Mxf}), for all $p^2$,  renormalized at a scale $\mu^2$, i.e. with $Z_2(\mu^2, \Lambda^2)$ such that $Z_R(\mu^2,\mu^2) = 1$.
If one multiplies both equations (\ref{Zxf}), (\ref{Mxf}) by $Z_R(\nu^2,\mu^2)$, it is clear that $Z_R(p^2,\nu^2)$ of Eq.\ (\ref{ZRnu}), and the original mass function $M(p^2)$ are solutions of the new equations, satisfying $Z_R(\nu^2,\nu^2)=1$, and with renormalization constant $Z_2(\nu^2,\Lambda^2)$ obeying Eq.\ (\ref{ZRnu}). The ability to construct the solutions renormalized at $\nu^2$ starting from the solutions renormalized at $\mu^2$, following the principles of multiplicative renormalization defined in Eq.\ (\ref{renormZ}), demonstrates that the novel truncation respects the multiplicative renormalizability of the solutions. Such a construction is not possible for Eqs.\ (\ref{ZxAb}), (\ref{MxAb}) of the bare Abelian approximation, hence showing its violation of multiplicative renormalizability. 

Also note that the solutions of the truncated equations (\ref{Zxf}), (\ref{Mxf}) satisfy the leading order resummed perturbative results. This feature was demonstrated earlier for the Abelian case \cite{Atkinson:1988mw, Atkinson:1990fj}, and even though the newly proposed equations differ from the former in their $Z$-dependence, their UV-limits are identical, as $Z_R(p^2) = 1$ to leading order in the Landau gauge, and hence, their mass functions have the same UV behavior. For systems where $Z_R(p^2) \ne 1$ the Abelian approximation does not reproduce the correct perturbative behavior for the quark functions, while truncations similar to Eqs.\ (\ref{UZ0}), (\ref{UM0}) still do. This has previously been shown for the gluon and ghost dressing functions in QCD \cite{Bloch:2001wz}, which obey similar equations. Also note that a truncation of the fermion equation in QED$_3$ \cite{Atkinson:1990fp}, constructed in order to reproduce the correct perturbative behavior, features a similar linearity in the dressing function $Z$ which, as was shown above, is sufficient to achieve multiplicative renormalizability. The treatment of QCD is, however, significantly different from QED, as the running coupling only enters the equations after  introducing appropriate ghost field corrections as proposed in Eqs.\  (\ref{UZ0}), (\ref{UM0}).

The fact that the simple truncation (\ref{UZ0}), (\ref{UM0}) satisfies these two important features of gauge theories is not a mere accident. Similarly to our work on the gluon equation in Ref.\ \cite{Bloch:2001wz}, the cancellations assumed in our Ansatz can be related to the cancellation of the quantum corrections coming from both the full quark-gluon vertex and the integral over the DSE kernel, in a way similar to Mandelstam's work on the gluon equation \cite{Mandelstam:1979xd}. Indeed, the quark-gluon vertex Slavnov-Taylor identity, which is a consequence of gauge invariance, is
\be\label{STIq1}
(p-k)^\mu \, \Gamma_{\mu}(p,k;p-k) = G((p-k)^2) \left( S_F^{-1}(k) - S_F^{-1}(p) \right) \,,
\ee
which shows that the bare vertex receives nonperturbative corrections proportional to $G/Z$. Although the Slavnov-Taylor identity does not constrain the transverse part of the vertex, multiplicative renormalizability requires the complete full vertex to have corrections of a similar nature. Furthermore, Mandelstam also showed in Ref.\ \cite{Mandelstam:1979xd} how perturbative loop corrections to the propagators in fact introduce a double $G/Z$ correction, and we herein assume that this is not just a perturbative feature, but rather a true nonperturbative one, engendered by the integrations over the full Green's functions in the self-energy integrals. Hence, in our analysis, one factor $Z/G$ is assumed to cancel the quantum corrections of the full vertex, and the other factor $Z/G$ to cancel the nonperturbative corrections generated by the integrals over the kernels. This nonperturbative Ansatz is consistent with perturbation theory at large momenta, hence yielding the correct resummed perturbative behavior of the quark functions, but moreover it preserves the multiplicative renormalizability of the solutions for any choice of renormalization point. Note that, even though the full vertex has not been constructed explicitly, it is genuinely \textit{nonperturbative} in our truncation -- however, the final equations contain bare vertices only, as a consequence of the non-trivial cancellations assumed in our Ansatz. 

The main aim of this study is to investigate the realization of dynamical chiral symmetry breaking in the proposed MR preserving truncation, where the kernel is explicitly driven by the running coupling. From the derivation of Eqs.\ (\ref{Zxf}), (\ref{Mxf}) we note that the running coupling in these equations is the true QCD running coupling, and not an effective coupling as was the case in the Abelian approximation, Eqs.\ (\ref{ZxAb}), (\ref{MxAb}). The running coupling connects the quark equation to the gauge sector, and in the next section we discuss the strong running coupling in more detail.

\section{The strong running coupling}

Formally the truncated quark equations (\ref{Zxf}), (\ref{Mxf}) have to be solved in a coupled system together with the equations for the gluon and ghost propagators presented in Ref.\ \cite{Bloch:2001wz}, such that the back reaction of the quarks on the gluon vacuum polarization be taken into account correctly in a truncation that preserves multiplicative renormalizability. However, as shown in Ref.\ \cite{Bloch:2001wz}, the pure gauge gluon-ghost system itself is not yet fully understood, and we can circumvent this complicated analysis, and postpone the study of the coupled equations to a later stage, by introducing a realistic model for the running coupling, thus decoupling the quark equations from the gluon-ghost system.

We will construct a model running coupling using information originating from perturbation theory, from Dyson-Schwinger studies of the coupled ghost-gluon system, and from recent lattice calculations.

\subsection*{Asymptotic behavior}

Renormalization group equation improved leading order perturbation theory yields the asymptotic ultraviolet behavior of the running coupling, including the quark contribution to the vacuum polarization:
\be\label{alpha-uv}
\alpha(q^2) \stackrel{\scriptscriptstyle{\text{UV}}}{\sim} \frac{4\pi}{\beta_0 \log(q^2/\LQCD^2)} \,,
\ee
where
\be\label{beta0}
\beta_0 = \frac{11 N_c - 2 N_f}{3} \,,
\ee
with $N_c$ colors and $N_f$ flavors, and where the leading order $\LQCD$ can be determined from some high energy experiment. Note that this leading order resummed behavior is also reproduced in the solutions of the nonperturbative DSEs, when computed using the truncation described in Ref.\ \cite{Bloch:2001wz}, which preserves multiplicative renormalizability.

From an analysis of the coupled ghost and gluon DSE system, von Smekal et al. \cite{vonSmekal:1998is} showed that, in a specific truncation scheme, the gluon and ghost propagators obey power laws in the infrared region, and the running coupling has an infrared fixed point. Furthermore, the leading infrared behavior of the gluon equation is completely determined by the ghost loop contribution to the vacuum polarization. These results were confirmed by studies of Atkinson and Bloch \cite{Atkinson:1998tu, Atkinson:1998zc} using different vertex Ans\"atze and angular approximations. 
However, the various truncations used in these studies all violate the principle of multiplicative renormalizability, and Ref.\ \cite{Bloch:2001wz} proposed a truncation where multiplicative renormalizability is respected, and which possesses the distinctive feature that all diagrams contribute to the leading infrared behavior of the gluon equation.
A recent investigation of the contributions of the two-loop diagrams to the gluon vacuum polarization \cite{Bloch:2001b} in this MR truncation reveals a relatively large infrared contribution of the four-gluon or \textit{squint} diagram which is essential, when combined with the ghost and gluon loop contributions \cite{Bloch:2001wz}, to ensure the existence of the propagator power laws, and of an infrared fixed point for the strong running coupling. This is a satisfying observation as the infrared power law solutions for both the gluon and the ghost propagators, and the infrared fixed point of the strong coupling are also supported by a number of recent lattice calculations \cite{Bonnet:2000kw, Langfeld:2001cz, Langfeld:2001a}.

\subsection*{The model}

Taking into account all this evidence, we herein build a model running coupling which has an infrared fixed point, and a constraint on its value can be derived from the ghost DSE only, as shown in Appendix \ref{App:alpha}, independently of the details of the gluon DSE. From Eq.\ (\ref{alpha0}) we compute that, for SU(3), the fixed point satisfies $2\pi/3 < \alpha_0 < 4\pi/3$, depending on the precise value of the exponent $\kappa$ of the propagator power laws (\ref{powlaw}), and combining with recent lattice results giving $\kappa \approx 0.5$ \cite{Bonnet:2000kw, Langfeld:2001cz, Langfeld:2001a}, we assume a preferred value $\alpha_0 = 5\pi/6 \approx 2.6$.

Even though we pinned down the leading asymptotic behavior of the running coupling at large and small momenta, we still need to know the behavior of the running coupling for all momenta in order to solve the quark equations (\ref{Zxf}), (\ref{Mxf}). To model the intermediate momentum region, where the coupling evolves from the infrared fixed point into the logarithmic tail, we use information collected from the study, Ref.\ \cite{Atkinson:1998tu}, of the coupled ghost-gluon system. The numerical results and the analytically calculated infrared asymptotic series in that study showed that the coupling remains very close to the infrared fixed point almost up to $q^2=\LQCD^2$, crosses the leading order perturbative curve, and thereafter drops quite rapidly to rejoin the perturbative curve, from above, around $q^2=10 \, \LQCD^2$. 

\begin{figure}{
\includegraphics[width=8cm]{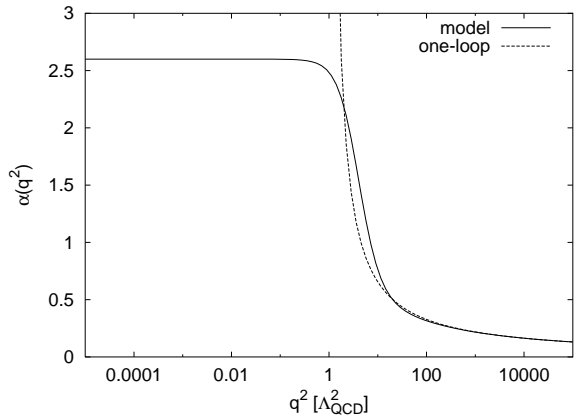}
\caption{\label{Fig:alpha}Running coupling $\alpha(q^2)$, as given by Eq.\ (\ref{alpha}) with $\alpha_0=2.6$ and $c_0=15$, and $q^2$ given in units $\LQCD^2$. The coupling has an infrared fixed point $\alpha_0$, and an ultraviolet behavior consistent with resummed leading order perturbation theory.}}
\end{figure}

We propose a functional expression for the running coupling which satisfies the various constraints mentioned above,  making use of rational polynomial approximations multiplying the required asymptotic behaviors. In its simplest form our approximation contains only one, dimensionless, parameter $c_0$ apart from the infrared fixed point $\alpha_0$ and the intrinsic scale $\LQCD$,
\be\label{alpha}
\alpha(q^2) \equiv \alpha(t \, \LQCD^2) = \frac{1}{c_0 + t^2}
\left[c_0 \, \alpha_0 + \frac{4\pi}{\beta_0}\left(\frac{1}{\log t}-\frac{1}{t-1}\right) t^2 \right] \,,
\ee
where $t=q^2/\LQCD^2$  \footnote{A more elaborate fit for $\alpha(q^2)$ would be\\ 
$\alpha(q^2) = \frac{1}{c_0 + c_1 t + t^2}
\left[(c_0 + a_1 t) \alpha_0 + \frac{4\pi}{\beta_0}\left(\frac{1}{\log t}-\frac{1}{t-1}\right) (t^2 +b_1 t) \right]$, with $t=q^2/\LQCD^2$.
}. The first term in the square brackets of Eq.\ (\ref{alpha}) is responsible for the infrared fixed point, while the term between round brackets yields the correct leading order logarithmic tail, where the simple pole at $\LQCD$ has been subtracted to make the coupling analytic for all spacelike momenta \cite{Shirkov:1997wi}. To match the intermediate region with the results of Ref.\ \cite{Atkinson:1998tu}, we choose  $c_0=15$, and we illustrate this model for the running coupling in Fig.\ \ref{Fig:alpha}.
Note that in contrast to the effective coupling often used in the application of DSE to hadron phenomenology, the running coupling (\ref{alpha}) does not have any infrared enhancement, and can be considered as a smoothened version of the concatenation of the infrared fixed point with the leading order logarithmic tail, where the smoothening is performed in a way suggested by previous studies of the gauge sector. The meaning and determination of the intrinsic QCD scale $\LQCD$ in the model (\ref{alpha}), will be discussed in the result section.

\section{Numerical results}

We now study how dynamical chiral symmetry breaking gets realized in Eqs.\ (\ref{Zxf}), (\ref{Mxf}) with kernels driven by $\alpha(q^2)$ of Eq.\ (\ref{alpha}). Note that $\alpha(q^2)$ possesses an intrinsic scale $\LQCD$, with which all momenta are scaled, and hence all momenta and masses in the quark equations can be expressed hereinafter in units of $\LQCD$. The equations are solved numerically for the chiral case, and the results, which satisfy MR, are compared with those computed in the MR-violating Abelian approximation. We investigate the sensitivity of the generated mass scale to the parameters of the model coupling, and discuss the determination of $\LQCD$ of the model. We also show the results of the quark equation in the massive case. 

The seeds of Eq.\ (\ref{Zxf}), (\ref{Mxf}) depend on the unknown renormalization constant $Z_2$ and bare mass $m_0$, and they are eliminated in the usual way by subtracting each equation at the renormalization point $\mu^2$ from that at $p^2$, and imposing the renormalization conditions $Z_R(\mu^2)=1$, $M(\mu^2)=m_\mu$. All integrals in the subtracted equations are finite, and the ultraviolet regulator can be taken to infinity. After introducing spherical coordinates, and performing two trivial angular integrations, we find
\ba\label{Zxsub}
\frac{1}{Z_R(x)} &=& 1 + \frac{C_F }{2\pi^2} \int_0^\infty dy \, 
\frac{1}{Z_R(y)}\frac{y}{y + M^2(y)}
\int_0^\pi d\theta \, \sin^2\theta \, \left [ \left ( \frac{\alpha(z)}{x}
\left [ \frac{3 \sqrt{yx}\cos\theta}{z} - \frac{2xy\sin^2\theta}{z^2}  \right ] \right ) - (x \leftrightarrow \mu^2 ) \right ] , \\
\label{Mxsub}
\frac{M(x)}{Z_R(x)} &=& m_\mu + \frac{3 \, C_F }{2\pi^2} \int_0^\infty dy \,
\frac{M(y)}{Z_R(y)} \frac{y}{y + M^2(y)}\,  \int_0^\pi d\theta \, \sin^2\theta \, \left[ \left ( \frac{\alpha(z)}{z} \right ) - (x \leftrightarrow \mu^2 ) \right]\,,
\ea
with $x=p^2$, $y=q^2$ and $z= x+y-2\sqrt{x y}\cos\theta$. Note that from now on the renormalization point dependence of $Z_R$ will be implicitly understood, $Z_R(x) \equiv Z_R(x,\mu^2)$.

\subsection*{Chiral case}

In the chiral case the mass equation (\ref{Mxf}) is homogeneous, as $Z_2 m_0 \equiv 0$, and no subtraction is needed for the mass equation -- Eq.\ (\ref{Mxsub}) is replaced by
\be
\label{Mxchir}
\frac{M(x)}{Z_R(x)} =  \frac{3 \, C_F }{2\pi^2} \int_0^\infty dy \,
\frac{M(y)}{Z_R(y)}\frac{y}{y + M^2(y)} \,  \int_0^\pi d\theta \, \sin^2\theta \, \frac{\alpha(z)}{z}  \,.
\ee

We substitute our model running coupling (\ref{alpha}) in the quark equations (\ref{Zxsub}), (\ref{Mxchir}) and compute the solutions of these equations using a numerical procedure first constructed to study dynamical chiral symmetry breaking in strong coupling QED \cite{Bloch:thesis}, and later also applied to solve the coupled gluon-ghost equations \cite{Atkinson:1998tu}. The method uses Chebyshev polynomials to approximate the unknown functions, and solves for the polynomial coefficients using the quadratically convergent Newton iterative method.

\begin{figure}{
\includegraphics[width=8cm]{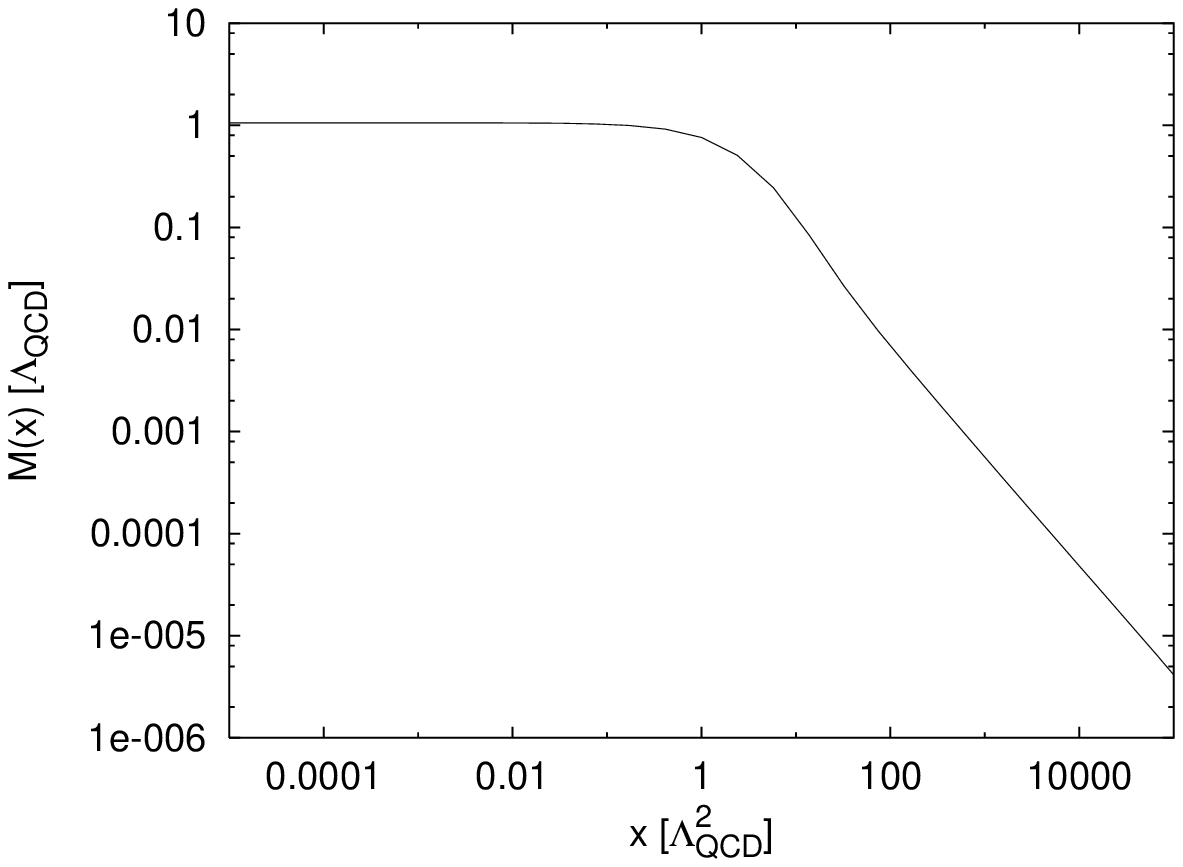}
\hfill\includegraphics[width=8cm]{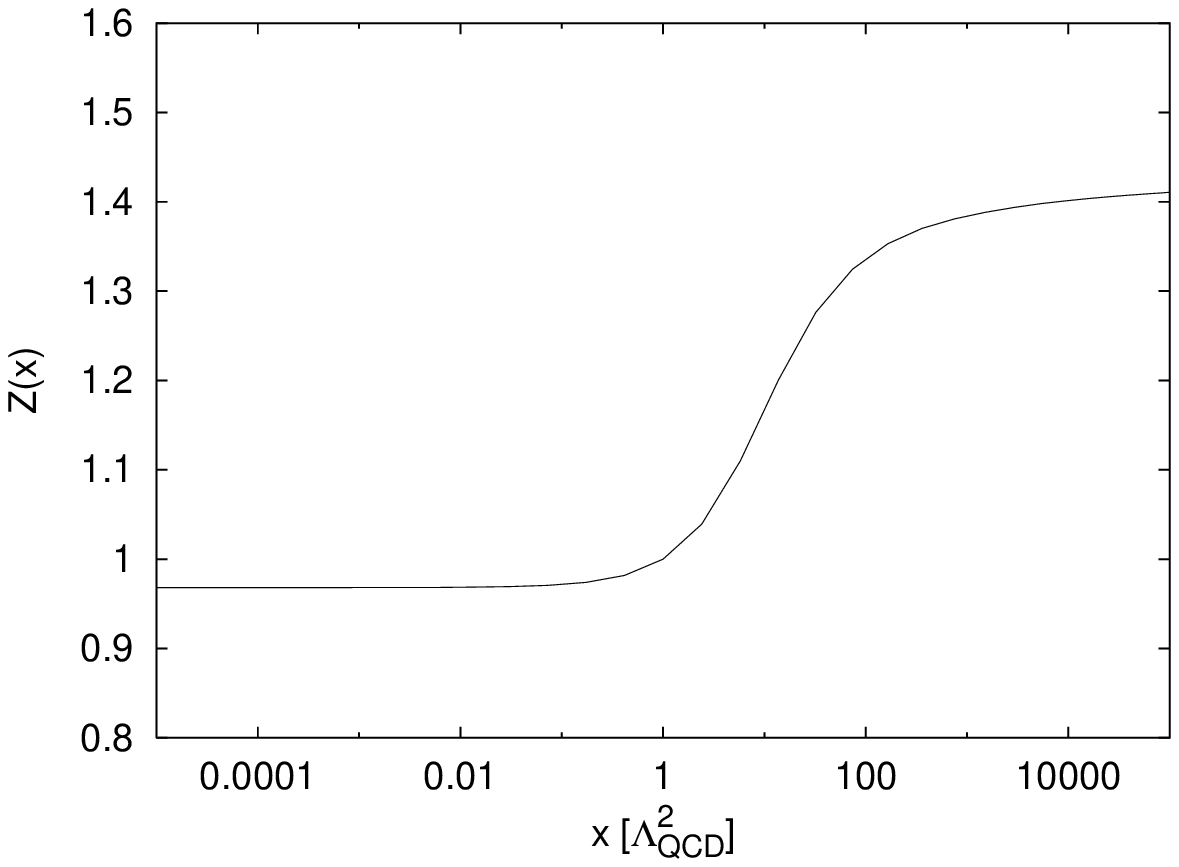}
\caption{\label{Fig:MZ}Solutions for the mass function $M(x)$ and quark dressing function $Z_R(x)$, from Eqs.\ (\ref{Zxsub}), (\ref{Mxchir}), renormalized at $\mu^2=\LQCD^2$, using $\alpha(x)$ given by Eq.\ (\ref{alpha}), with $\alpha_0=2.6$ and $c_0=15$, in the chiral case. Preservation of multiplicative renormalizability means that a different choice of renormalization point will merely multiply $Z_R(x)$ by a constant factor, and leave $M(x)$ unchanged.}
}\end{figure}

The evolution of $M(x)$ and $Z_R(x)$ as a function of momentum is shown in Fig.\ \ref{Fig:MZ}, for $10^{-5} \le x/\LQCD^2 \le 10^5$, and renormalization point $\mu^2 = \LQCD^2$.
The mass function is expressed in units of $\LQCD$, which sets the scale in the running coupling (\ref{alpha}). Note that, in our MR preserving truncation, a different choice of renormalization point will merely multiply the whole function $Z_R(x)$ by a finite, constant factor over the complete momentum range, and will leave the mass function $M(x)$ unchanged. The strength of the dynamical chiral symmetry breaking can be characterized by the value of the mass function at the origin:
\be\label{M0}
M(0) \approx 1.057 \, \LQCD \,.
\ee
The generated quark mass is of the order of the extension of the infrared plateau of the coupling, and the absence of fine-tuning in the coupling, to obtain such a mass scale, points to a certain naturalness in the MR truncation. 

In the chiral case the ultraviolet behavior of the dynamical mass function satisfies the well-known asymptotic formula \cite{Lane:1974he, Politzer:1976tv}: 
\be\label{pertsolMch}
M(x) \sim \frac{2 \pi^2 \gamma_m}{3} \frac{-\langle\bar{q} q\rangle^0}{x \, [\frac{1}{2}\log(x/\Lambda_{QCD}^2)]^{1-\gamma_m}} \,,
\ee
with $\gamma_m = 12/(33-2N_f)$. A large momentum fit yields the value for the renormalization-point independent vacuum quark condensate, which is also representative for the strength of dynamical chiral symmetry breaking:
\be\label{qq0}
-\langle\bar{q} q \rangle^0 \approx (0.70 \, \LQCD)^3 \,.
\ee

\subsection*{Abelian approximation}

An important part of our analysis is to evaluate the influence of the multiplicatively renormalizable approach on the dynamically generated mass scale. The only difference between the MR Eqs.\ (\ref{Zxf}), (\ref{Mxf}) and Eqs.\ (\ref{ZxAb}), (\ref{MxAb}) of the bare vertex, Abelian approximation, is the factor $Z_R(y)$ which appears in the denominator of the integration kernels in the former, while it is found in their numerator in the latter, in a way violating MR. Using our model running coupling (\ref{alpha}) as effective coupling in the Abelian approximation, we find a dynamically generated mass that strongly depends on the renormalization point, as can be seen in Table \ref{Tab:nonMR}. The value which is usually quoted in Abelian studies is that of the unrenormalized case, for which the dynamical mass is $M(0) \approx 0.326 \, \LQCD$, about three times smaller than in the herein presented MR preserving truncation. Our MR truncation clearly disagrees with the use of a bare quark-gluon vertex, and the implicit dressing of the vertex assumed in the MR truncation is essential. The violation of MR, and hence the renormalization point dependence of the generated mass scale makes the bare vertex Abelian truncation ill defined.

We furthermore observe that the mass scale generated in the MR truncation depends very little on the momentum evolution of $Z$. To show this, it suffices to decouple the mass equation (\ref{Mxchir}) from the $Z$-equation (\ref{Zxsub}) by forcing $Z_R(x) \equiv 1$ for all $x$ in the mass equation. The generated mass is then $M(0)=1.082 \, \LQCD$, only 2\% 
off the value of the coupled system of equations.

\subsection*{Model sensitivity}

We now study the sensitivity of the solutions of the MR truncated quark equations to the features of the model (\ref{alpha}) for the running coupling.

First, we examine which part of the kinematical regime of the running coupling (\ref{alpha}) mainly contributes to the dynamical mass in the MR truncation. We substitute the solutions to Eqs.\ (\ref{Zxsub}), (\ref{Mxchir}), shown in Fig.\ \ref{Fig:MZ}, in the right hand side of Eq.\ (\ref{Mxchir}), and calculate the mass integral for varying values of the upper integration limit $\Lambda_{\text{UV}}^2$. The results are given in Table \ref{Tab:UV}, and we find that slightly more than half of the dynamical mass is generated by the infrared fixed point region of the coupling, $p^2 < \LQCD^2$. Most of the remaining mass is generated in the transition region, $\LQCD^2 < p^2 < 10 \, \LQCD^2$, while the logarithmic tail only contributes about 1\%. 
Hence, the dynamical mass is mainly generated in the nonperturbative region of the coupling, and is virtually independent of its precise ultraviolet behavior. It is therefore not essential, in the context of dynamical chiral symmetry breaking, that the ultraviolet behavior of the coupling (\ref{alpha}) only reproduces the RGE improved leading order perturbative results.

\begin{figure}[t]
\parbox{8cm}{\includegraphics[width=8cm]{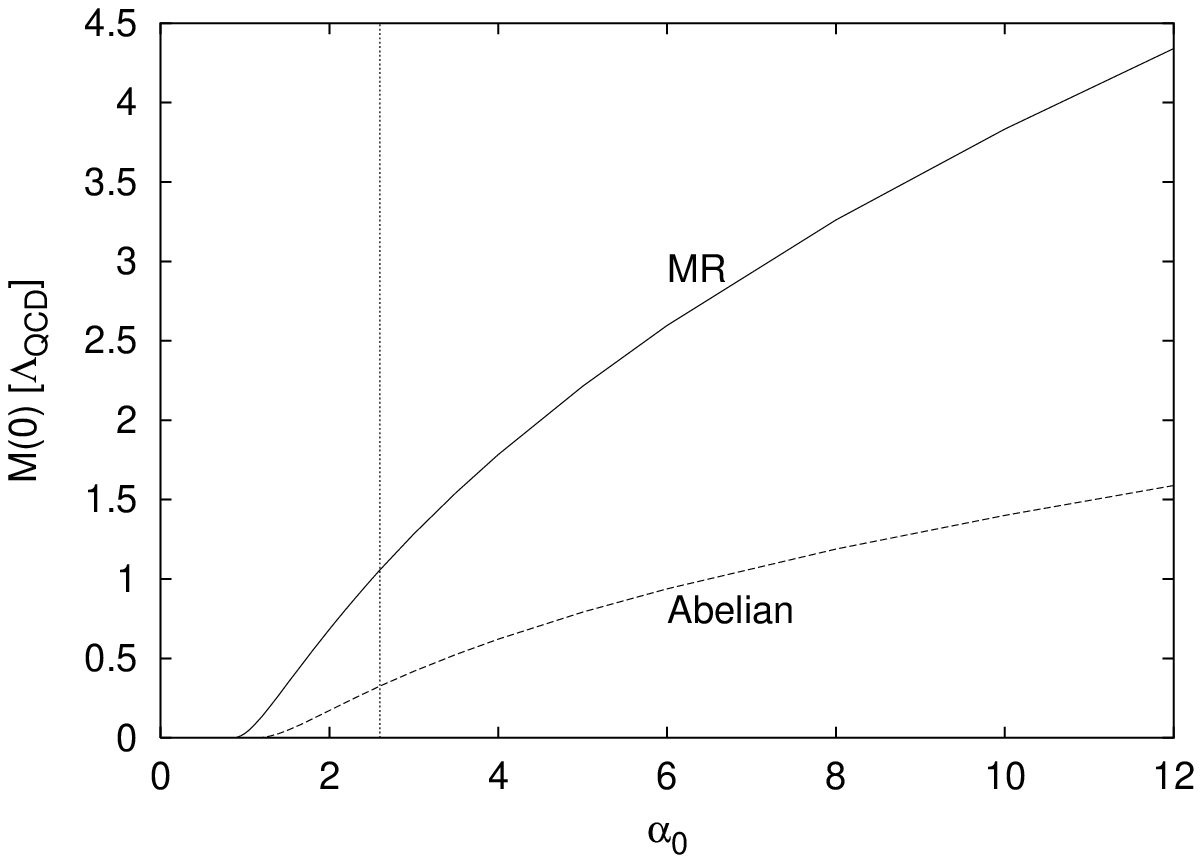}
\caption{\label{Fig:Mvalpha}Variation of the generated mass $M(0)$, in units of $\LQCD$, with varying $\alpha_0$ in Eq.\ (\ref{alpha}), and $c_0=15$, for our MR truncation scheme, and in the Abelian approximation. The vertical line shows the preferred value $\alpha_0=2.6$.}}
\hfill
\parbox{8cm}{\includegraphics[width=8cm]{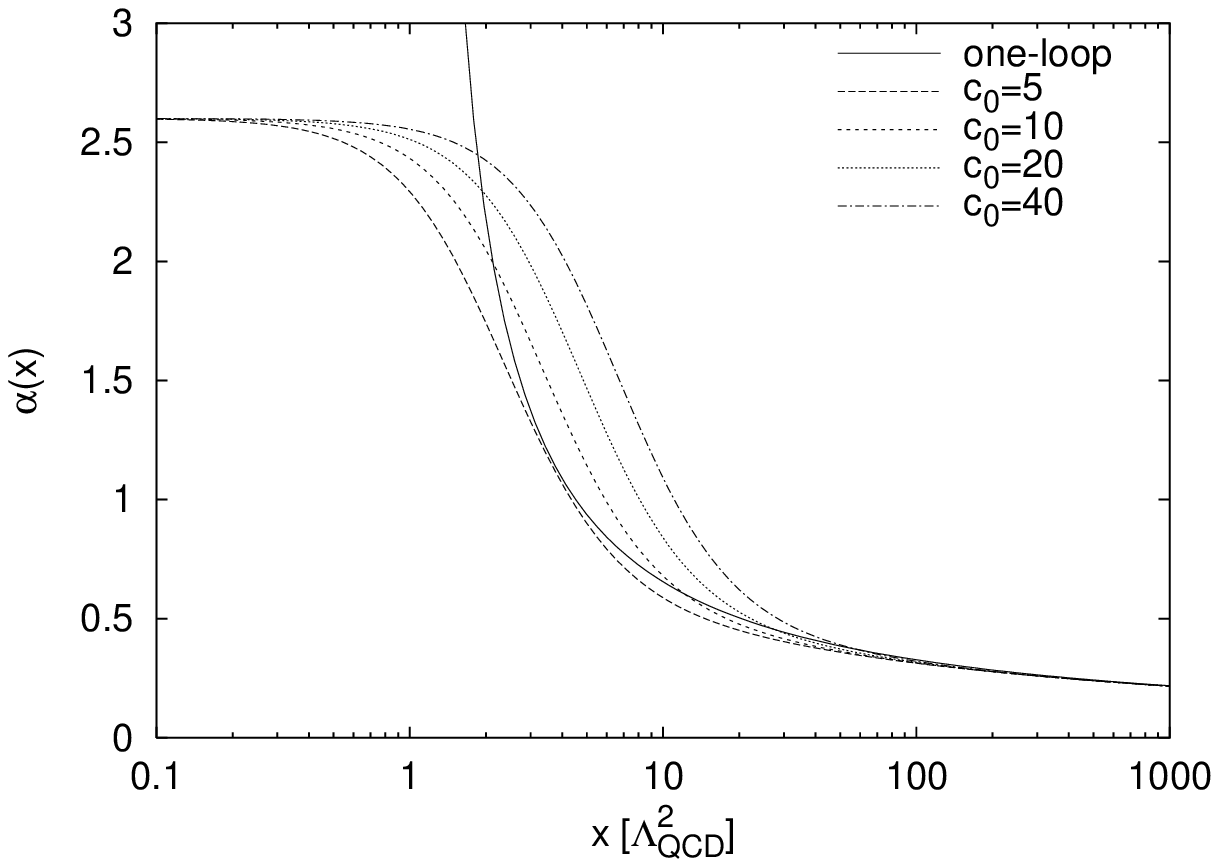}
\caption{\label{Fig:alphac0}Running coupling $\alpha(x)$, given by Eq.\ (\ref{alpha}) with $\alpha_0=2.6$, for a range of $c_0$'s. To focus on the influence of $c_0$ on the intermediate momentum region, the plotted momentum range is taken $0.1 \le x/\LQCD^2 \le 1000$.}}
\end{figure}

Next, we investigate the sensitivity of the results to the parameters of our model coupling (\ref{alpha}). As shown in Appendix \ref{App:alpha}, the value of the fixed point depends on the details of the infrared analysis of the gluon-ghost DSE system, with bounds $2\pi/3 < \alpha_0 < 4\pi/3$, and preferred value $5\pi/6$. In Fig.\ \ref{Fig:Mvalpha} we show how the generated mass changes as we vary the value of the fixed point over a slightly wider range, $1 \le \alpha_0 \le 12$, and some of the main values are also tabulated in Table \ref{Tab:alpha0}. 
\begin{table}[t]
\parbox{7cm}{\begin{tabular}{|c|@{~~}c@{~~}|@{~~}r@{~~}|}
\hline
& $\mu^2$ & $M(0)$ \\
\hline
&\\[-3ex]
MR & any\,\, & 1.057 \\
\hline
\hline
&\\[-3ex]
Abelian
& unren & 0.326 \\
& $10^5$ &  0.327 \\
& $10^3$ &  0.355 \\
& $1$ & 1.262 \\
& $10^{-3}$ & 1.398 \\
& $10^{-5}$ & 1.398 \\
\hline
\end{tabular}
\caption{\label{Tab:nonMR}Dynamically generated mass $M(0)$, in units $\LQCD$, versus renormalization point $\mu^2$ (in $\LQCD^2$) in MR-violating Abelian approximation, compared with MR solution.}}
\hspace{10mm}
\parbox{7cm}{\begin{tabular}{|@{~~}c@{~~}|@{~~}r@{~~}|}
\hline
$\Lambda_{\text{UV}}^2$ & $M(0)$ \\
\hline
&\\[-3ex]
0.1 &  0.076 \\
1 &    0.562 \\
10 &   1.044 \\
100 &  1.057 \\
$10^3$ & 1.057 \\
\hline
\end{tabular}
\caption{\label{Tab:UV}Dynamically generated mass $M(0)$, in units $\LQCD$, versus upper integration limit $\Lambda_{UV}^2$ (in $\LQCD^2$) in Eq.\ (\ref{Mxchir}).}}
\end{table}
\begin{table}[b]
\parbox{7cm}{\begin{tabular}{|@{~~}r@{~~}|@{~~}r@{~~}|}
\hline
$\alpha_0$ & $M(0)$ \\
\hline
&\\[-3ex]
1.0 & 0.028 \\
1.8 & 0.548 \\
2.6 & 1.057 \\
3.4 & 1.494 \\
4.2 & 1.874 \\
\hline
\end{tabular}
\caption{\label{Tab:alpha0}Sensitivity of the dynamically generated mass $M(0)$, in units $\LQCD$, to changes in the infrared fixed point of the coupling, $\alpha_0$ of Eq.\ (\ref{alpha}), with $c_0=15$.}}
\hspace{10mm}
\parbox{7cm}{\begin{tabular}{|@{~~}r@{~~}|@{~~}r@{~~}|}
\hline
$c_0$ & $M(0)$ \\
\hline
&\\[-3ex]
5 & 0.824 \\
10 & 0.964 \\
15 & 1.057 \\
20 &  1.129 \\
40 &   1.325 \\
\hline
\end{tabular}
\caption{\label{Tab:c0}Sensitivity of the dynamically generated mass $M(0)$, in units $\LQCD$, to changes in the parameter $c_0$ of Eq.\ (\ref{alpha}), with $\alpha_0=2.6$.}}
\end{table}
In the figure we also plot the corresponding results for the unrenormalized, bare vertex, Abelian case discussed earlier. As could be expected there is a critical value $\alpha_0^{\text{crit}} \approx 0.90$ (for $c_0=15$) below which chiral symmetry is not spontaneously broken. However, the fixed point value dictated by the QCD equations is well above this value, and the quarks do acquire an appropriately sized dynamical mass in the MR truncation.

When constructing the model coupling (\ref{alpha}), the shape of its drop in the intermediate region, $\LQCD^2 < p^2 < 10 \, \LQCD^2$, where the coupling flows from the infrared asymptotic behavior into its perturbative logarithmic tail, was inferred from the numerical results of Ref.\ \cite{Atkinson:1998tu}. Even though we are confident that the qualitative features of this study will remain valid outside the scope of its specific truncations (angular approximation, violation of MR, disregard of quark loop and two-loop diagrams), more will be learned about the intermediate regime when solving the coupled quark-gluon-ghost set of equations in an MR preserving truncation, as we plan to do in future work. However, the sensitivity of the generated mass scale with respect to a change in intermediate behavior can already be studied by varying the parameter $c_0$ of our model coupling (\ref{alpha}). The behavior of the running coupling, as $c_0$ is varied, is illustrated in Fig.\ \ref{Fig:alphac0}, and the dynamically generated mass is tabulated in Table \ref{Tab:c0}, for selected values of $c_0$. The parameter $c_0$ is related to the position of the transition region, where the running coupling drops from its fixed point to rejoin the perturbative curve, and the dynamical mass gradually increases with increasing $c_0$.  

Note that another approach yielding a finite infrared value of the running coupling can be derived from an analytic continuation of the perturbative coupling as shown in Ref.\ \cite{Shirkov:1997wi}. In contrast to the model used above, the coupling derived in that approach has no infrared plateau, and is exactly given by expression (\ref{alpha}) with $c_0=0$, yielding a finite infrared value  $\alpha(0)=1.51$, for $N_f=4$. This running coupling does not yield enough integration strength in the gap equation, as the computed dynamical fermion mass $M(0) = 0.036 \, \LQCD$  is clearly too small for hadronic applications.

\subsection*{Determining $\LQCD$}

As the running coupling $\alpha(q^2)$ of Eq.\ (\ref{alpha}) is a function of $q^2/\LQCD^2$, all mass scales in the quark equations are expressed in units $\LQCD$, and the value of $\LQCD$ has to be determined by matching theory with experiment.

Determining $\LQCD$ of our model by matching $\alpha$ in the perturbative region is not a reliable procedure, as the perturbative behavior of our model coupling (\ref{alpha}) only agrees with perturbation theory up to resummed leading order, and small changes in $\alpha$ in the logarithmic region will have a big impact on the value of $\LQCD$, and on all related hadronic masses, as can be inferred from Eqs.\ (\ref{M0}), (\ref{qq0}). 

In perturbation theory $\LQCD$ corresponds to the position of the pole in the logarithmic expansion of the running coupling, and depends on the renormalization scheme, and on the mass threshold matching procedure  \cite{Hinchliffe:2000rz}. In the $\overline{\text{MS}}$ scheme its five flavor value  determined from experiment is $\LMSb^{(5)}=208^{+25}_{-23} \, \MeV$ \cite{Hinchliffe:2000rz}. However, $\LQCD$ changes as flavor thresholds are crossed, and the running of the perturbative coupling should be calculated using the appropriate $\LQCD^{(N_f)}$. Taking into account mass thresholds, one finds $\LMSb^{(4)}=309^{+33}_{-30} \, \MeV$ below the b-quark mass, and $\LMSb^{(3)}=377^{+35}_{-33} \, \MeV$ below the c-quark threshold  (as long as perturbation theory remains valid). 

The nonperturbative model (\ref{alpha}) has no perturbative pole, and there $\LQCD$ is uniquely related to the evolution of the coupling (\ref{alpha}), and corresponds to the scale where the coupling starts its descent off the infrared plateau towards the perturbative regime (see Fig.\ \ref{Fig:alpha}). As this definition of $\LQCD$ is at variance with the perturbative definition, a straightforward identification of its value with the perturbative values is not possible. Our aim herein is to investigate nonperturbative effects, like dynamical chiral symmetry breaking, hence $\LQCD$ of our model should be determined by matching some nonperturbative quantity related to the scale of hadronic physics.

If we take $\LQCD = 330 \, \MeV$ in the model coupling (\ref{alpha}), the MR truncation generates a dynamical quark mass $M(0) \approx 350 \, \MeV$ and a chiral condensate $-\langle\bar{q} q\rangle^0 \approx (230 \, \MeV)^3$, according to Eqs.\ (\ref{M0}), (\ref{qq0}), which are of the right order of magnitude for hadron phenomenology. Moreover, this value of $\LQCD$ is consistent with the above mentioned perturbative values, taking into account their different definitions, as the coupling grows strong in the same momentum window.

Furthermore, for such a value of $\LQCD$, our model coupling (\ref{alpha}), whose ultraviolet behavior is only modelled up to leading order in perturbation theory, yields a perturbative value $\alpha(M_Z) = 0.134$ ($M_Z=91.187\,\GeV$). Particle physics phenomenology gives a world-average $\alpha_s(M_Z) = 0.118$ for the strong coupling \cite{Hinchliffe:2000rz}, and the leading order contribution to this value, computed from perturbation theory, is $\alpha_s^{\text{1-loop}}(M_Z) = 0.135$, including mass-threshold effects \cite{Ellis:1996qj}. We can thus conclude that our model coupling produces results which are consistent with both nonperturbative and perturbative physics.

\begin{figure}{
\includegraphics[width=8cm]{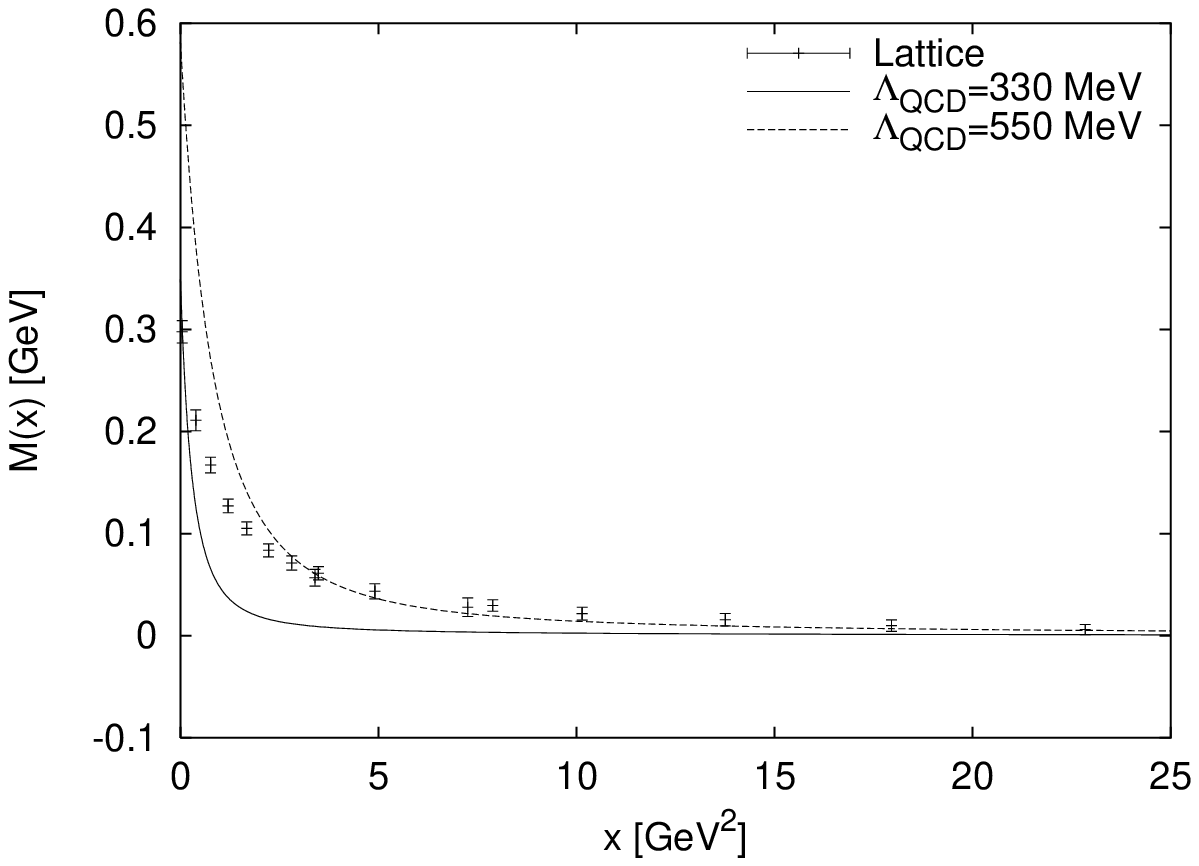}
\hfill\includegraphics[width=8cm]{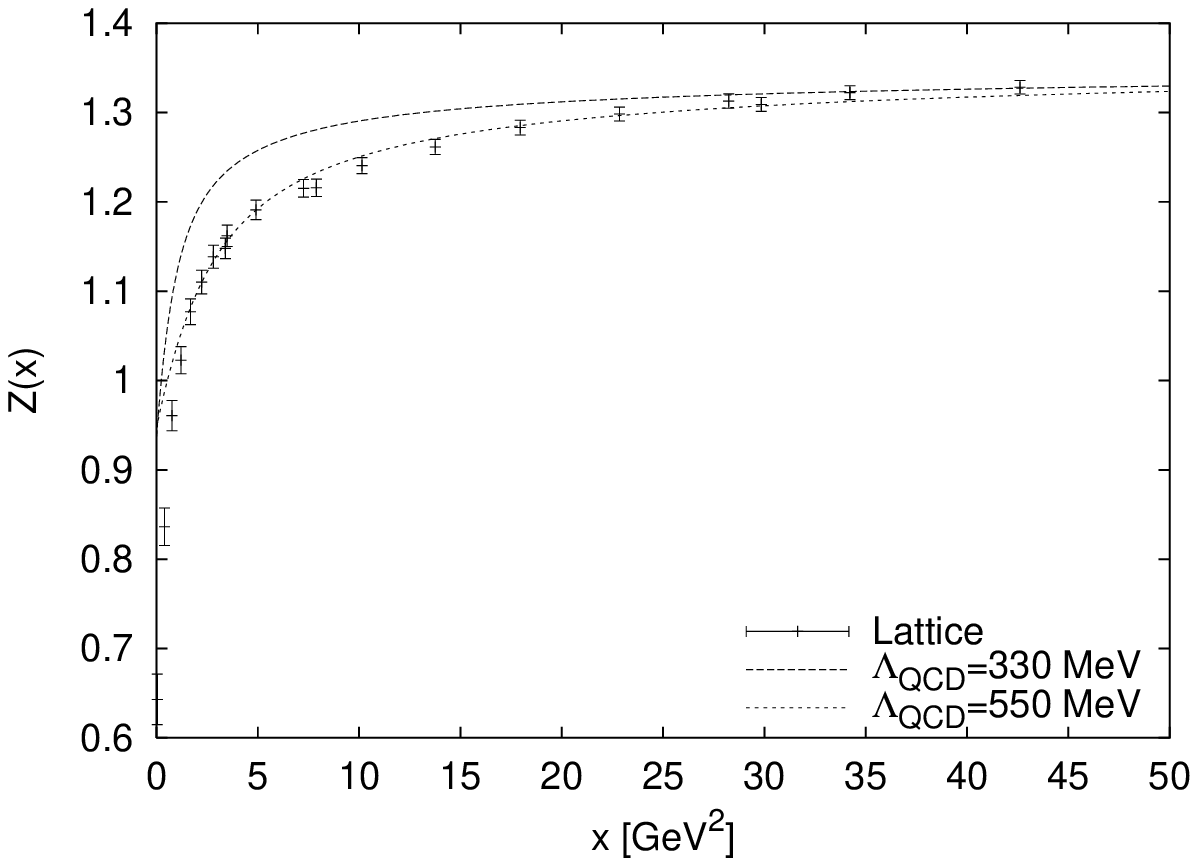}
\caption{\label{Fig:MZcomp}Comparison of the Dyson-Schwinger solutions for the mass function $M(x)$ and quark dressing function $Z_R(x)$ in the chiral case, with the lattice results of Ref.\ \cite{Bonnet:2002ih} (chiral extrapolation). The Dyson-Schwinger solutions are computed from Eqs.\ (\ref{Zxsub}), (\ref{Mxchir}), using $\alpha(x)$ given by Eq.\ (\ref{alpha}), with $\alpha_0=2.6$,  $c_0=15$, and with $\LQCD=330 \, \MeV$ and $550 \, \MeV$ (see text). The lattice results of Ref.\ \cite{Bonnet:2002ih} are plotted as a function of the kinematic lattice momentum.}
}\end{figure}

The numerical results obtained herein can also be compared with some of the latest lattice results on the quark propagator in the Landau gauge (with $N_f=0$) \cite{Bonnet:2002ih}. Fig.\ \ref{Fig:MZcomp} shows the lattice data, and the DSE results with the phenomenologically preferred value $\LQCD=330\,\MeV$, and also with $\LQCD=550\,\MeV$, which yields a better agreement with the lattice points in the intermediate momentum regime ($3\,\GeV^2<x<50\,\GeV^2$). Note that below $3\,\GeV^2$ the lattice results have a slower increase of the mass function, and a steeper decrease of the dressing function. It is this discrepancy at small momenta that affects the value of $\LQCD$ when matching the DSE-results with the lattice data. Possible origins are lattice artefacts, vertex truncations, or inaccuracies in the modeling of the coupling (\ref{alpha}) in the intermediate momentum regime. As the size of the dynamically generated mass is crucial for hadronic physics, this disparity should be investigated further. 

\subsection*{Massive case}

\begin{figure}
\includegraphics[width=8cm]{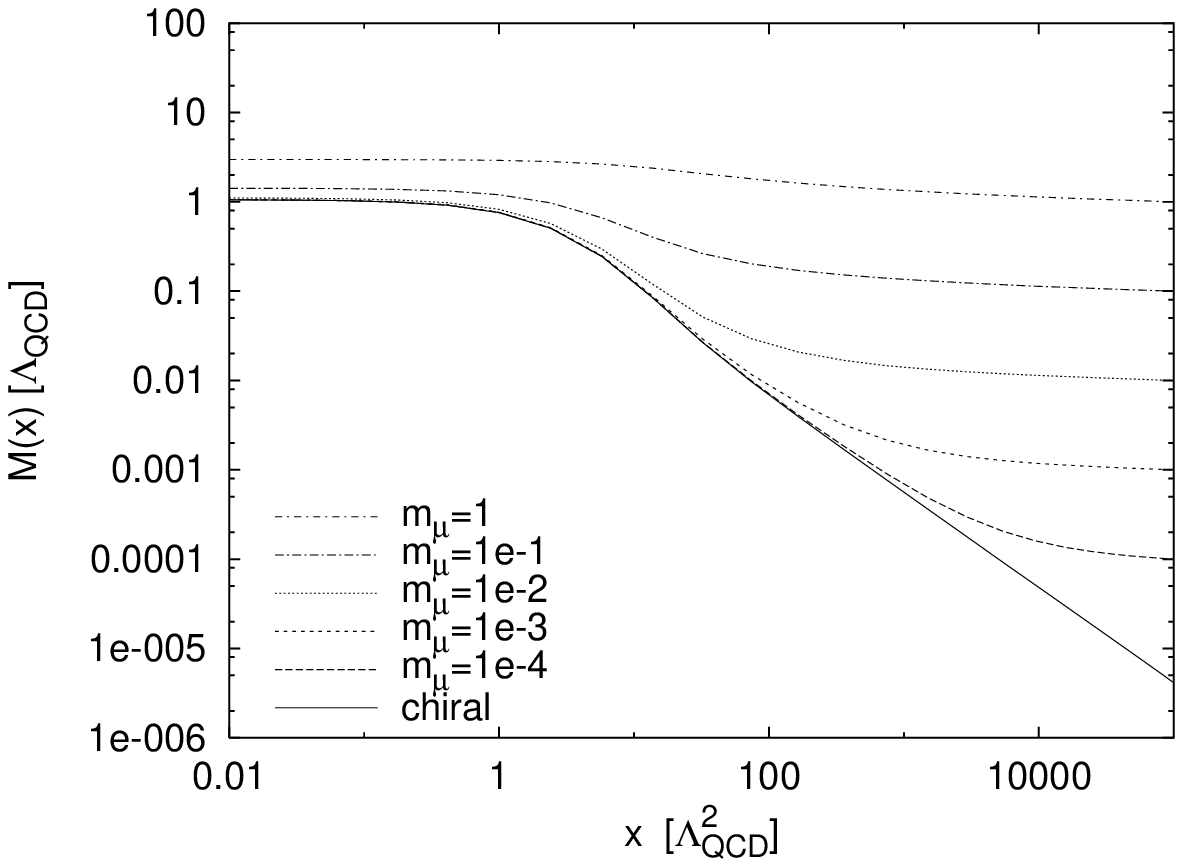}
\hfill\includegraphics[width=8cm]{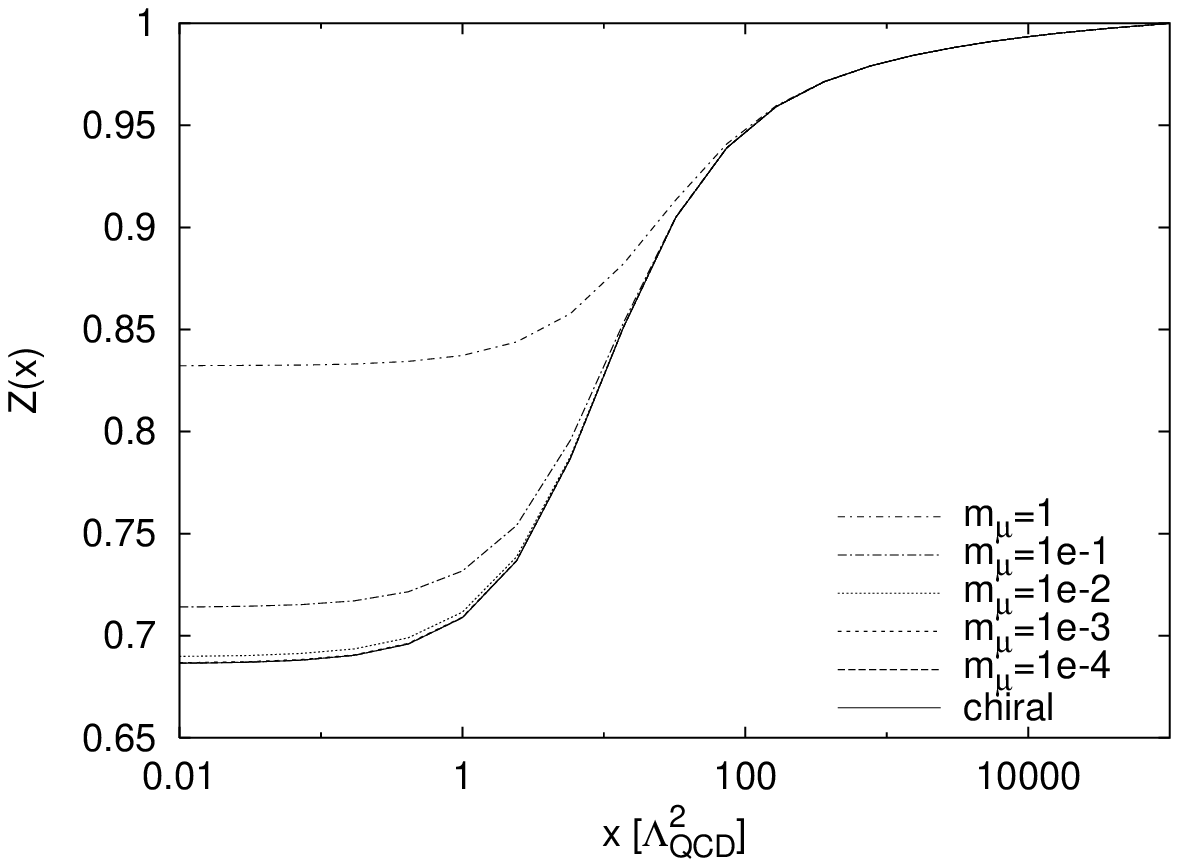}
\caption{\label{Fig:MZr}Mass function and quark dressing function solutions from Eqs.\ (\ref{Zxsub}), (\ref{Mxsub}), using $\alpha(x)$ given by Eq.\ (\ref{alpha}), with $\alpha_0=2.6$ and $c_0=15$, for various values of renormalized quark mass $m_\mu$, with renormalization point $\mu^2/\LQCD^2=10^5$. Note the influence of the quark mass on the onset of the logarithmic tail in $M(x)$.}
\end{figure}

Next we extend our analysis to the massive case, $Z_2 m_0 \neq 0$ in the mass equation (\ref{Mxf}), or in practice $m_\mu > m_\mu^{\text{chiral}}$ in Eq.\ (\ref{Mxsub}), where $m_\mu^{\text{chiral}} = M^{\text{chiral}}(\mu^2)$. In Fig.\ \ref{Fig:MZr} we plot the mass and quark dressing functions for varying renormalized mass, $m_\mu/\LQCD=0.0001, 0.001, 0.01, 0.1, 1$, together with the chiral case (with renormalization point $\mu^2/\LQCD^2 = 10^5$). 
In the massive case the logarithmic behavior satisfies the RGE improved perturbative result, 
\be\label{mpert}
M(p^2) = m_\mu \left[ \frac{\alpha(p^2)}{\alpha_\mu} \right]^{\gamma_m} 
= \frac{\hat m}{\left[ \log (p^2/\LQCD^2) \right]^{\gamma_m}} \,,
\ee
where $\hat m = m_\mu \left[ \log(\mu^2/\LQCD^2)\right]^{\gamma_m}$, and $\gamma_m = 12/(33 - 2N_f)$, which is the correct anomalous dimension of the mass function. 
We observe that the log-tail of the mass function sets in for ever smaller momenta as the renormalized mass is tuned above the chiral value. Note again that the preservation of multiplicative renormalizability allows us to choose an arbitrary renormalization point, without altering the physical content of the results: changing the renormalization point merely multiplies $Z_R(x)$ by a constant factor at all momenta, and leaves the mass function $M(x)$ unchanged.

\subsection*{Discussion}

As shown before, the Abelian approximation, which violates MR, requires a strong infrared enhancement of the effective coupling to achieve a strong enough breaking of the chiral symmetry, and such an enhancement seems inconsistent with the characteristics of the gauge sector. 
A first attempt to extend the DSE studies beyond the Abelian approximation by solving a truncated set of DSEs for the quark, gluon and ghost propagators simultaneously \cite{Ahlig:thesis} finds a dynamically generated chiral condensate of $(122 \, \MeV)^3$ (with $N_f=4$), which is about a factor two too small for phenomenological purposes, apparently confirming the problem to find the necessary integration strength in the quark equation. 
The study uses a QCD analog \cite{Alkofer:1998mw} of the quenched QED Curtis-Pennington vertex \cite{Curtis:1990zs} in an attempt to preserve gauge invariance and MR. This is however not sufficient:  the vertex renormalization constant $Z_{1f}$ is constructed such that the solutions to the equations have the correct leading order perturbative behavior, but it is not consistent with the chosen nonperturbative vertex Ansatz, and its renormalization point dependence violates multiplicative renormalizability.

In Ref.\ \cite{Kizilersu:2000qd} a method was presented to solve the renormalized quark equations in an MR truncation of quenched QED. There the explicit construction of $Z_{1f}$ is circumvented by eliminating the renormalization constants from the equations. It is however not clear if and how the method could be extended to QCD. Moreover, in contrast to our MR truncation, their method requires an explicit construction of a nonperturbative vertex Ansatz satisfying the Ward-Takahashi identity and conditions of MR, and the two truncation schemes are not equivalent. 

The MR truncation scheme presented in this paper leads to solutions that satisfy the principles of multiplicative renormalizability in an elegant way, without explicit constructions of full vertices or renormalization constants, while naturally producing a large enough dynamical breaking of chiral symmetry in QCD.
In a next step, the MR truncation derived for the gluon-ghost system \cite{Bloch:2001wz} should be combined with that proposed herein, to solve the coupled set of DSEs for quark, gluon and ghost propagators, thus avoiding the explicit construction of a model running coupling.
Consequently, an MR preserving truncation scheme for the Bethe-Salpeter equation, describing hadronic bound states, should be constructed in a way consistent with the quark equation derived herein, which respects the Goldstone boson nature of the pion.

\section{Conclusions}

We have reformulated the coupled set of continuum equations for the quark dressing function and dynamical mass function in QCD, such that, in the Landau gauge, all renormalization constants are eliminated, and the multiplicative renormalizability of the quark dressing function, and renormalization point invariance of the mass function are manifest. The kernels of the equations explicitly depend on the momentum evolution of the strong running coupling. 
The new formulation allowed us to construct a truncation which preserves these properties, and also satisfies the leading order renormalization group equation improved perturbative results. This is achieved by assuming non-trivial cancellations involving the full quark-gluon vertex in the self-energy loop, such that no explicit construction of the full vertex and vertex renormalization constant are needed. 

In order to perform numerical calculations of the dynamically generated mass scale, without solving the coupled quark-gluon-ghost DSE system, we decoupled the quark equations from the gluon and ghost equations by introducing a model running coupling. This coupling satisfies the leading order ultraviolet behavior known from perturbative QCD, and its nonperturbative behavior was constrained using the most recent results of gluon-ghost DSE and lattice studies. The coupling has an infrared fixed point ($\alpha_0 \approx 2.6$), and its behavior in the intermediate momentum region reproduces the qualitative features of the numerical DSE studies.

The integral equations for the quark propagator are solved numerically for the chiral and massive cases, and the fermion mass and chiral condensate generated dynamically in the MR preserving truncation is of the order of the extension of the infrared plateau of the coupling. The generated scale is about three times larger than what is found using the MR-violating, bare vertex, Abelian approximation, and of the right order of magnitude to perform hadron phenomenology, without  infrared enhancement in the strong coupling.

\begin{acknowledgments}
 I thank  R.~Alkofer, D.~Atkinson, C.~Fischer, K.~Langfeld, S.~M.~Schmidt, and P.~Watson for useful comments and discussions. This work was funded by Deutsche Forschungsgemeinschaft under project no.~SCHM 1342/3-1.
\end{acknowledgments}

\appendix

\section{Infrared behavior of the strong running coupling}
\label{App:alpha}

We briefly show how the value of the infrared fixed point $\alpha_0$ of the strong coupling should be determined from the coupled Dyson-Schwinger equations for the gluon and ghost propagators. The contributions of the various diagrams in the gluon equation are however not yet fully understood, and we describe how the ghost equation alone can give us tight bounds on the value of $\alpha_0$.

The DSE for the gluon propagator is
\be
\label{SD-Gl-full}
\left[D_{\mu\nu}(p)\right]^{-1} = \left[D^0_{\mu\nu}(p)\right]^{-1}
- \pi_{\mu\nu}^{gh}(p) - \pi_{\mu\nu}^{gl}(p) - \pi_{\mu\nu}^{3g}(p) - \pi_{\mu\nu}^{4g}(p) - \pi_{\mu\nu}^{tad}(p) - \pi_{\mu\nu}^{q}(p)  \,,
\ee
where $D^0_{\mu\nu}$ and $D_{\mu\nu}$ are the bare and full gluon propagators, and the vacuum polarization includes contributions from the ghost loop, gluon loop, three-gluon diagram, four-gluon diagram, tadpole diagram and quark loop. The DSE for the ghost propagator, in Euclidean space, is
\be
\label{SD-Gh}
\left[\Delta(p)\right]^{-1} = \left[\Delta^0(p)\right]^{-1} 
- N_c g_0^2 \int \frac{d^4 q}{(2\pi)^4} \, G^0_\mu(p,q) \Delta(q) G_\nu(q,p) D^{\mu\nu}(r)  \,,
\ee
where $N_c$ is the number of colors, $g_0$ is the bare coupling, $\Delta^0$ and $\Delta$ are the bare and full ghost propagators, $G_\mu^0$ and $G_\mu$ the bare and full ghost-gluon vertices, and $r=p-q$.
The general expressions for the full gluon and ghost propagators, defining the gluon and ghost dressing functions $F(p^2)$ and $G(p^2)$, were given in Eqs.\ (\ref{glprop}), (\ref{ghprop}).

As mentioned in the main body of this paper, the coupled equations Eqs.\ (\ref{SD-Gl-full}), (\ref{SD-Gh}) have been solved simultaneously in a number of previous studies using various truncations, and there is convincing evidence that $F$ and $G$ obey power laws in the infrared, when all the diagrams in the gluon equation are taken into account properly:
\be\label{powlaw}
F_R(x,\mu^2) \sim x^{2\kappa} \quad , \qquad  G_R(x,\mu^2) \sim x^{-\kappa} \,,
\ee
with $x=p^2$, and where $F_R$ and $G_R$ are the renormalized gluon and ghost dressing functions defined in Eq.\ (\ref{renorm}). 

Because of Eq.\ (\ref{rgi2}), these power laws lead to an infrared fixed point for the running coupling,
\be\label{nu}
\alpha_0 = \lim_{x \to 0} \alpha(\mu^2) \, F_R(x,\mu^2) \, G_R^2(x,\mu^2)  \to \text{constant} \,.
\ee

As explained in detail in Ref.\ \cite{Bloch:2001wz}, the ghost and gluon equations each typically yield a relation between the infrared fixed point $\alpha_0$ and the leading infrared exponent $\kappa$:  
\be\label{condit}
N_c \, \alpha_0  = \frac{1}{\chi_{gh}(\kappa)} \quad , \qquad 
N_c \, \alpha_0 = \frac{1}{\chi_{gl}(\kappa)} \,,
\ee
when substituting the power laws (\ref{powlaw}) in the coupled equations (\ref{SD-Gl-full}), (\ref{SD-Gh}), and equating the coefficients of the leading power of $x$ for $x \to 0$ on both sides of the equations. The functions $\chi_{gh}(\kappa)$, $\chi_{gl}(\kappa)$ are computed from the vacuum polarization integrals, and a consistent infrared power solution requires the gluon and ghost expressions in Eq.\ (\ref{condit}) to be satisfied simultaneously:
\be\label{consist}
\chi_{gh}(\kappa) = \chi_{gl}(\kappa) \,.
\ee
The solution of this equation yields the value of the leading infrared exponent $\kappa$, and the corresponding $\alpha_0$ can then be computed from Eq.\ (\ref{condit}).

In a recent study \cite{Bloch:2001wz} we have shown that the gluon function $\chi_{gl}(\kappa)$ is a complicated object as it requires the computation of the nonperturbative loop integrals for all the gluon vacuum polarization diagrams in the infrared. However, \textit{if} the power laws are valid, \textit{then} a consistent value of $\kappa$ exists, and the ghost equation can, on its own, provide very useful information. Indeed, the first identity of Eq.\ (\ref{condit}) relates the value of the infrared fixed point $\alpha_0$ of the strong coupling to the infrared exponent $\kappa$ of the dressing functions (\ref{powlaw}), without needing additional information from the gluon identity. Therefore, we now further analyze the infrared behavior of the ghost equation (\ref{SD-Gh}) in the bare ghost-gluon vertex approximation, and compute $\chi_{gh}$.

In the Landau gauge, the bare vertex approximation of Eq.\ (\ref{SD-Gh}) for the renormalized ghost dressing function becomes,
\be
\label{SD6}
\frac{1}{G_R(x,\mu^2)} = \tilde Z_3(\mu^2,\Lambda^2)
- \frac{N_c \,\alpha_\mu}{4 \pi^3}  \intfour \, T_0(x,y,z)\, G_R(y,\mu^2) \, F_R(z,\mu^2) \,,
\ee
where we introduced the full propagators (\ref{glprop}), (\ref{ghprop}), the renormalized quantities (\ref{renormghgl}), (\ref{renorm}), used  $\tilde Z_1 \equiv 1$ (in the Landau gauge), set $x=p^2$, $y=q^2$, $z=r^2$, and
\be
\label{Txyz0}
T_0(x,y,z) =
- \left({\frac{{x}}{y}} -
      2 + {\frac{{y}}{x}}\right){\frac{1}{4\,z^2}} +
  \left({\frac{1}{y}} + {\frac{1}{x}}\right){\frac{1}{2\,z}} - {\frac{1}{4\,x\,y}} \,.
\ee

After substitution of the power laws (\ref{powlaw}) in Eq.\ (\ref{SD6}), the right hand side yields a sum of integrals of the form
\be\label{integ-xyz}
\intfour \, x^\alpha \, y^\beta \, z^\gamma \,,
\ee
with $\alpha+\beta+\gamma = \kappa - 2$. In Ref.\ \cite{Atkinson:1998zc} these integrals were solved using spherical coordinates, and the results were expressed in terms of generalized hypergeometric functions. A more concise, equivalent expression can be derived \cite{Lerche:2001} by noting that
integrals of type (\ref{integ-xyz}) are typically encountered when applying dimensional regularization in perturbative calculations, and they are readily computed by introducing Feynman parameters, yielding \cite{Peskin:1995ev}
\be
I(a,b) = \int \frac{d^4 q}{(2\pi)^4}\, \frac{1}{y^{a} z^{b}} 
= \frac{1}{16\pi^2} \frac{\Gamma(2-a)\Gamma(2-b)\Gamma(a+b-2)}{\Gamma(a)\Gamma(b)\Gamma(4-a-b)} \, x^{2-a-b}\,.
\ee

Both sides of Eq.\ (\ref{SD6}) yield a leading infrared power $x^\kappa$, and equating their coefficients gives
\be\label{alpha0}
\alpha_0 = \frac{2\pi}{3 N_c} 
\frac{\Gamma(3-2\kappa)\Gamma(3+\kappa)\Gamma(1+\kappa)}{\Gamma^2(2-\kappa)\Gamma(2\kappa)} \,,
\ee
where the Gamma function recurrence relation was used repeatedly to bring the expression in its simplest form.

\begin{figure}
{\includegraphics[width=8cm]{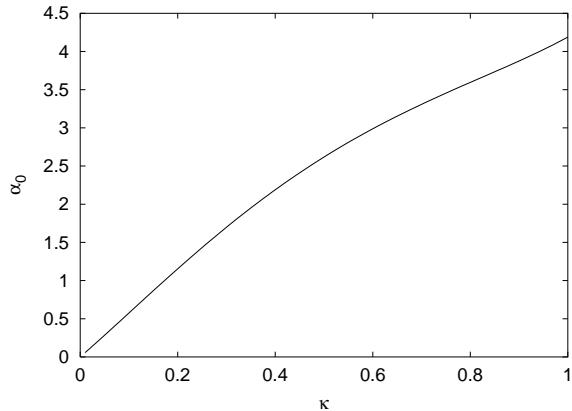}
\caption{\label{Fig:alphak}Fixed point $\alpha_0$ of the strong running coupling versus infrared exponent $\kappa$, as computed from the bare vertex approximation to the ghost equation for SU(3).}}
\end{figure}

Expression (\ref{alpha0}) gives the relation between the infrared fixed point of the coupling and the infrared exponent $\kappa$ imposed by the ghost equation, and we illustrate this dependence for SU(3) in Fig.\ \ref{Fig:alphak}. As mentioned before, the gluon equation \textit{will} be consistent with this expression for some value of $\kappa$ if the infrared power laws for the propagators are valid. Previous DSE studies \cite{vonSmekal:1998is, Atkinson:1998tu, Atkinson:1998zc, Lerche:2001} yielded values for $\kappa$ between 0.4 and 1.0, depending on the truncation, and recent lattice calculations \cite{Bonnet:2000kw, Langfeld:2001cz} seem to indicate that $\kappa$ is close to 0.5. Although the exact value of $\kappa$ depends on the consistency condition (\ref{consist}) between the ghost and gluon DSEs, we see from Fig.\ \ref{Fig:alphak} that the value of $\alpha_0$ is tightly bound, and will lie in the interval $[2\pi/3,4\pi/3]$ for SU(3), and the preferred value used herein is $\alpha_0(\kappa=0.5) = 5\pi/6 \approx 2.6$. Note that the large values for the fixed point, $\alpha_0 \approx 10-12$, mentioned in some previous studies \cite{vonSmekal:1998is, Atkinson:1998tu}, have been shown to be artefacts of angular approximations in the integration kernels \cite{Atkinson:1998zc}. 


\begin{thebibliography}{10}

\bibitem{'tHooft:1971fh}
G.~'t~Hooft,
\newblock Nucl. Phys. {\bf B33}, 173 (1971).

\bibitem{Politzer:1973fx}
H.~D. Politzer,
\newblock Phys. Rev. Lett. {\bf 30}, 1346 (1973).

\bibitem{Roberts:1994dr}
C.~D. Roberts and A.~G. Williams,
\newblock Prog. Part. Nucl. Phys. {\bf 33}, 477 (1994), [hep-ph/9403224].

\bibitem{Roberts:2000aa}
C.~D. Roberts and S.~M. Schmidt,
\newblock Prog. Part. Nucl. Phys. {\bf 45S1}, 1 (2000), [nucl-th/0005064].

\bibitem{Brown:1988bm}
N.~Brown and M.~R. Pennington,
\newblock Phys. Rev. {\bf D38}, 2266 (1988).

\bibitem{Frank:1996uk}
M.~R. Frank and C.~D. Roberts,
\newblock Phys. Rev. {\bf C53}, 390 (1996), [hep-ph/9508225].

\bibitem{Maris:1997tm}
P.~Maris and C.~D. Roberts,
\newblock Phys. Rev. {\bf C56}, 3369 (1997), [nucl-th/9708029].

\bibitem{Hawes:1994ef}
F.~T. Hawes, C.~D. Roberts and A.~G. Williams,
\newblock Phys. Rev. {\bf D49}, 4683 (1994), [hep-ph/9309263].

\bibitem{Hawes:1998cw}
F.~T. Hawes, P.~Maris and C.~D. Roberts,
\newblock Phys. Lett. {\bf B440}, 353 (1998), [nucl-th/9807056].

\bibitem{Jain:1993qh}
P.~Jain and H.~J. Munczek,
\newblock Phys. Rev. {\bf D48}, 5403 (1993), [hep-ph/9307221].

\bibitem{Atkinson:1988mw}
D.~Atkinson and P.~W. Johnson,
\newblock Phys. Rev. {\bf D37}, 2296 (1988).

\bibitem{Roberts:1990mj}
C.~D. Roberts and B.~H.~J. McKellar,
\newblock Phys. Rev. {\bf D41}, 672 (1990).

\bibitem{Alkofer:2000wg}
R.~Alkofer and L.~von Smekal,
\newblock Phys. Rept. {\bf 353}, 281 (2001), [hep-ph/0007355].

\bibitem{Mandelstam:1979xd}
S.~Mandelstam,
\newblock Phys. Rev. {\bf D20}, 3223 (1979).

\bibitem{Atkinson:1981er}
D.~Atkinson, J.~K. Drohm, P.~W. Johnson and K.~Stam,
\newblock J. Math. Phys. {\bf 22}, 2704 (1981).

\bibitem{Atkinson:1982ah}
D.~Atkinson, P.~W. Johnson and K.~Stam,
\newblock J. Math. Phys. {\bf 23}, 1917 (1982).

\bibitem{Brown:1989bn}
N.~Brown and M.~R. Pennington,
\newblock Phys. Rev. {\bf D39}, 2723 (1989).

\bibitem{vonSmekal:1998is}
L.~von Smekal, A.~Hauck and R.~Alkofer,
\newblock Ann. Phys. {\bf 267}, 1 (1998), [hep-ph/9707327].

\bibitem{Atkinson:1998tu}
D.~Atkinson and J.~C.~R. Bloch,
\newblock Phys. Rev. {\bf D58}, 094036 (1998), [hep-ph/9712459].

\bibitem{Atkinson:1998zc}
D.~Atkinson and J.~C.~R. Bloch,
\newblock Mod. Phys. Lett. {\bf A13}, 1055 (1998), [hep-ph/9802239].

\bibitem{Bonnet:2000kw}
F.~D.~R. Bonnet, P.~O. Bowman, D.~B. Leinweber and A.~G. Williams,
\newblock Phys. Rev. {\bf D62}, 051501 (2000), [hep-lat/0002020].

\bibitem{Langfeld:2001cz}
K.~Langfeld, H.~Reinhardt and J.~Gattnar,
\newblock Nucl. Phys. {\bf B} (2001, in press), [hep-ph/0107141].

\bibitem{Langfeld:2001a}
J.~C.~R. Bloch, A.~Cucchieri, K.~Langfeld and T.~Mendes,
\newblock in preparation.

\bibitem{Ahlig:thesis}
S.~Ahlig,
\newblock \protect{Ph.D. thesis, Universit\"at T\"ubingen}, url:
  http://w210.ub.uni-tuebingen.de/dbt/volltexte/2001/280/.

\bibitem{Bloch:2001wz}
J.~C.~R. Bloch,
\newblock Phys. Rev. {\bf D64}, 116011 (2001), [hep-ph/0106031].

\bibitem{Bloch:2001b}
J.~C.~R. Bloch,
\newblock in preparation.

\bibitem{Collins:1984xc}
J.~C. Collins,
\newblock {\em Renormalization} (Univ. Pr., Cambridge, Uk, 1984).

\bibitem{Taylor:1971ff}
J.~C. Taylor,
\newblock Nucl. Phys. {\bf B33}, 436 (1971).

\bibitem{Atkinson:1990fj}
D.~Atkinson and P.~W. Johnson,
\newblock Phys. Rev. {\bf D41}, 1661 (1990).

\bibitem{Atkinson:1990fp}
D.~Atkinson, P.~W. Johnson and P.~Maris,
\newblock Phys. Rev. {\bf D42}, 602 (1990).

\bibitem{Shirkov:1997wi}
D.~V. Shirkov and I.~L. Solovtsov,
\newblock Phys. Rev. Lett. {\bf 79}, 1209 (1997), [hep-ph/9704333].

\bibitem{Bloch:thesis}
J.~C.~R. Bloch,
\newblock Ph.D. thesis, Universtity of Durham.

\bibitem{Lane:1974he}
K.~D. Lane,
\newblock Phys. Rev. {\bf D10}, 2605 (1974).

\bibitem{Politzer:1976tv}
H.~D. Politzer,
\newblock Nucl. Phys. {\bf B117}, 397 (1976).

\bibitem{Hinchliffe:2000rz}
I.~Hinchliffe,
\newblock Eur. Phys. J. {\bf C15}, 85 (2000).

\bibitem{Ellis:1996qj}
R.~K. Ellis, W.~J. Stirling and B.~R. Webber,
\newblock Cambridge Monogr. Part. Phys. Nucl. Phys. Cosmol. {\bf 8}, 1 (1996).

\bibitem{Bonnet:2002ih}
CSSM Lattice, F.~D.~R. Bonnet, P.~O. Bowman, D.~B. Leinweber, A.~G. Williams
  and J.-b. Zhang,
\newblock hep-lat/0202003.

\bibitem{Alkofer:1998mw}
R.~Alkofer, S.~Ahlig and L.~von Smekal,
\newblock Fizika {\bf B8}, 277 (1999), [hep-ph/9901322].

\bibitem{Curtis:1990zs}
D.~C. Curtis and M.~R. Pennington,
\newblock Phys. Rev. {\bf D42}, 4165 (1990).

\bibitem{Kizilersu:2000qd}
A.~Kizilersu, A.~W. Schreiber and A.~G. Williams,
\newblock Phys. Lett. {\bf B499}, 261 (2001), [hep-th/0010161].

\bibitem{Lerche:2001}
C.~Lerche,
\newblock \protect{Diploma thesis, Universit\"at Erlangen}.

\bibitem{Peskin:1995ev}
M.~E. Peskin and D.~V. Schroeder,
\newblock {\em An introduction to quantum field theory} (Addison-Wesley,
  Reading, USA, 1995).

\end{thebibliography}

\end{document}